\documentclass[sn-standardnature,iicol]{sn-jnl}

\newcommand{\upmu}{{\usefont{U}{eur}{m}{n}\symbol{"16}}}

\usepackage[compact]{titlesec}
\titleformat{\section}{\normalfont\normalsize\bfseries}{0pt}{}{}

\jyear{2021}%

\begin{document}

\title[Article Title]{Observing high-k magnons with Mie-resonance-enhanced Brillouin light scattering}


\author*[1]{\fnm{Ond\v{r}ej} \sur{Wojewoda}}\email{ondrej.wojewoda@ceitec.vutbr.cz}

\author[1,2]{\fnm{Filip} \sur{Ligmajer}}\email{filip.ligmajer@ceitec.vutbr.cz}

\author[1,2]{\fnm{Martin} \sur{Hrtoň}}\email{martin.hrton@vutbr.cz}

\author[2]{\fnm{Jan} \sur{Kl\'{i}ma}}\email{jan.klima4@vutbr.cz}

\author[1]{\fnm{Meena} \sur{Dhankhar}}\email{dhankhar.meena13@gmail.com}

\author[2]{\fnm{Krist\'{y}na} \sur{Dav\'{i}dkov\'{a}}}\email{kristyna.davidkova@vut.cz}

\author[1]{\fnm{Michal} \sur{Sta\v{n}o}}\email{michal.stano@ceitec.vutbr.cz}

\author[1]{\fnm{Jakub} \sur{Holobr\'{a}dek}}\email{holobradek@vutbr.cz}

\author[1,2]{\fnm{Jakub} \sur{Zl\'{a}mal}}\email{zlamal@fme.vutbr.cz}

\author[1,2]{\fnm{Tom\'{a}\v{s}} \sur{\v{S}ikola}}\email{sikola@fme.vutbr.cz}

\author*[1,2]{\fnm{Michal} \sur{Urb\'{a}nek}}\email{michal.urbanek@ceitec.vutbr.cz}

\affil[1]{\orgdiv{CEITEC BUT}, \orgname{Brno University of Technology}, \orgaddress{\street{Purky\v{n}ova 123}, \city{Brno}, \postcode{612 00}, \country{Czech Republic}}}

\affil[2]{\orgdiv{Faculty of Mechanical Engineering, Institute of Physical Engineering}, \orgname{Brno University of Technology}, \orgaddress{\street{Technick\'{a} 2}, \city{Brno}, \postcode{616 69}, \country{Czech Republic}}}


\abstract{Magnonics is a prospective beyond CMOS technology which uses magnons, the quanta of spin waves, for low-power information processing. Many magnonic concepts and devices were recently demonstrated at macro- and microscale, and now these concepts need to be realized at nanoscale. Brillouin light scattering spectroscopy and microscopy (BLS) has become a standard technique for spin wave visualization and characterization, and enabled many pioneering magnonic experiments. However, due to its fundamental limit in maximum a detectable magnon momentum, the conventional BLS cannot be used to detect nanoscale spin waves. Here we show that optically induced Mie resonances in dielectric nanoparticles can be used to extend the range of accessible spin wave wavevectors beyond the BLS fundamental limit. The method is universal and can be used in many magnonic experiments dealing with thermally excited as well as coherently excited high-momentum, short-wavelength spin waves. This discovery significantly extends the usability and relevance of the BLS technique for nanoscale magnonic research.
}





\maketitle
Advances in Brillouin light scattering microscopy (BLS) \cite{Sebastian2015, Madami2012, Wang2020}, and spectroscopy \cite{Kargar2021}  significantly helped magnonics to become one of the most promising candidates for "beyond CMOS" technology \cite{IRDS:BC2021, Chumak2022}. For device miniaturization, a shift towards short-wavelength, high-momentum magnons is necessary. Until now, the BLS techniques fell short here due to their fundamental limit in maximum detectable magnon momentum \cite{Chumak2022, Liu2018, Wintz2016, Pirro2021, Baumgaertl2020}. This limit is given by the law of conservation of momentum in the Stokes process: $\boldsymbol{k}_\mathrm{i}=\boldsymbol{k}_\mathrm{s}+\boldsymbol{k}_\mathrm{mag}$, where $\boldsymbol{k}_\mathrm{i}$ and $\boldsymbol{k}_\mathrm{s}$ are $k$-vectors of the incident and scattered light and $\boldsymbol{k}_\mathrm{mag}$ is the $k$-vector of the magnon on which the light is being scattered. It means that in a typical BLS experiment in back-scattering geometry, the maximal detectable $k$-vector of spin waves equals twice the $k$-vector of the incident light. For the  laser wavelength $\lambda_\mathrm{i} = 532$\,nm, for example, the maximum $k$-vector that can be theoretically detected is $k^\mathrm{max}_\mathrm{mag} = 23.6\,\mathrm{rad}$\,\upmu m$^{-1}$, which corresponds to a minimum spin-wave wavelength $\lambda^\mathrm{min}_\mathrm{mag} = \lambda_\mathrm{i}/2 = 266\,\mathrm{nm}$ \cite{Sebastian2015, Madami2012}. Taking an inspiration from tip- and surface-enhanced Raman scattering spectroscopy \cite{Han2021, Zhang2013, Nie1997}, nanosized apertures or other plasmonic structures made of metals have been used to locally enhance the electromagnetic field and increase the range of the accessible $k$-vectors \cite{Uteglov2007, Jersch2010, Freeman2020}. Unfortunately, the efficiency of the plasmonic approach is severely limited by high optical losses in metallic structures which makes it unsuitable for convenient magnon measurements (see \cite{Freeman2020} and Extended Data Fig. \ref{Extfig-Plasmon}). However, recent advances in nanophotonics suggest that plasmonic structures made of metals can be substituted by structures made of dielectric materials. Such dielectric nanoresonators have an advantage in reduced dissipative losses and associated heating, while their high refractive index still enables comparatively strong light confinement \cite{Xiao2021, Genevet2017, Alessandri2016, Caldarola2015, Yesilkoy2019, Tseng2020, dorrah2021}.

\begin{figure}
    \centering
    \includegraphics{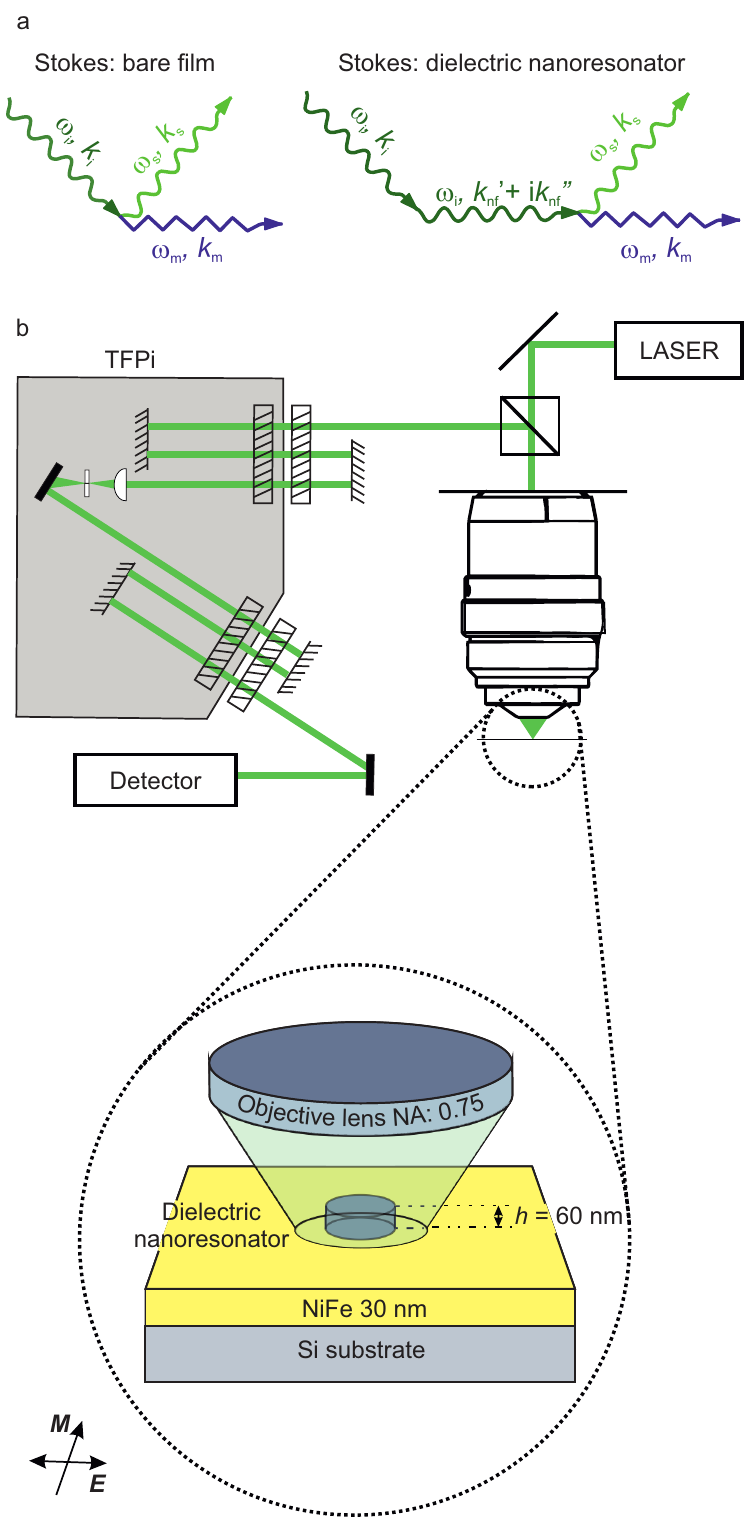}
    \caption{\textbf{Geometry of the experiment.} \textbf{a}, Comparison of the conventional and Mie resonance enhanced (Stokes) BLS processes. Dielectric nanoresonator converts an incoming photon into a photon with complex momentum. The real part of the converted photon can be larger than the momentum of the incident free-space light. This way, the limit of conventional BLS can be overcome, and high-$k$ magnons can be optically detected. \textbf{b}, Schematics of our \upmu-BLS setup. Laser light (532\,nm) is focused onto a silicon disk by a high numerical aperture objective lens. The disk is placed on the top of a permalloy layer, in which the spin waves are probed. The inelastically back-scattered light is then analyzed in the tandem Fabry-Perot interferometer (TFPi) and collected with a single-photon detector. }
    \label{fig1}
\end{figure}

Here we show that optically induced Mie resonances in dielectric nanoparticles can be used in magnonic BLS experiments to increase the magnon signal and to enhance the range of accessible $k$-vectors beyond the fundamental limit of conventional BLS microscopy setup (\upmu-BLS) \cite{Sebastian2015}. We use simple silicon disks, which support Mie resonances \cite{Evlyukhin2011, Staude2017, kuznetsov2016, koshelev2020} with strong and localized electric fields (hot spots). When the incident light with  momentum $k_\mathrm{i}$ is restricted to the sub-diffraction hot spots, its momentum becomes complex $k_\mathrm{nf} = k_\mathrm{nf}'+ik_\mathrm{nf}''$   and thus its real part $k_\mathrm{nf}'$ can be larger than the momentum of the free-space light \cite{Bezeres2013} (see Fig. \ref{fig1}a). This way, the fundamental limit of BLS in maximum detectable magnon momentum can be overcome.

In our experiments, we have investigated spin waves in a 30\,nm thick permalloy film, on top of which we fabricated arrays of 60\,nm thick silicon disks and the diameters ranging from 100\,nm to 1.5\,\upmu m. The sample was measured on a standard \upmu-BLS, using a microscope objective lens with $\mathrm{NA}=0.75$ to illuminate it by $\lambda_i = 532\,$nm coherent laser light (Fig. \ref{fig1}b). The scattered photons were collected by the same objective lens and guided to a tandem Fabry-Perot interferometer (TFPi). Prior to the measurement on the silicon disk, we verified the $k$-vector detection limit of our setup on a bare permalloy film. In contrast to the theoretically predicted value of $k^\mathrm{max} = \mathrm{NA}\cdot 23.6 \,\mathrm{rad}$\,\upmu m$^{-1} = 17.7\,\mathrm{rad}$\,\upmu m$^{-1}$, the estimated value for our setup turned out to be $k^\mathrm{max} = 10\,\mathrm{rad}$\,\upmu m$^{-1}$, in agreement with the experimentally measured laser spot size (440\,nm, see Extended Data Fig. \ref{Extfig-beamShape}). A more detailed description of the sample fabrication and the \upmu-BLS setup is given in the Methods section. 

\begin{figure}
    \centering
    \includegraphics{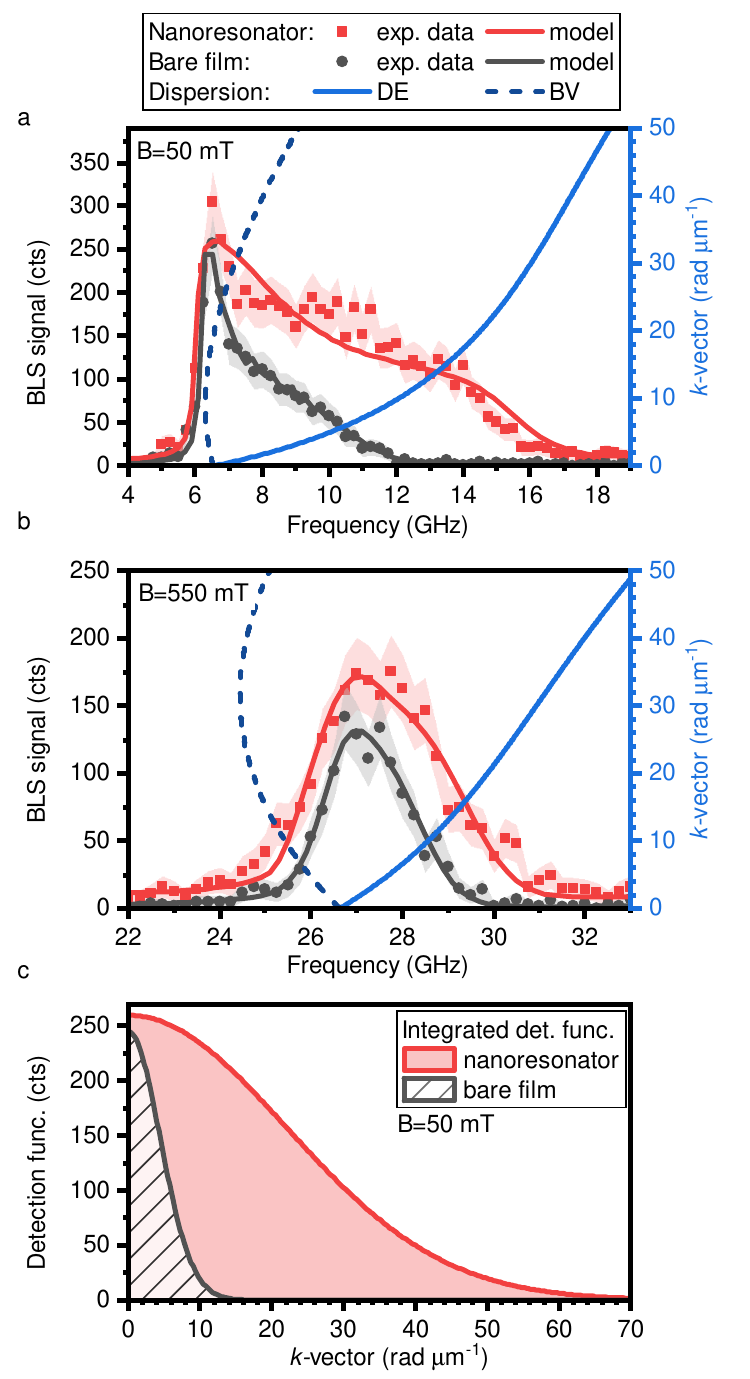}
    \caption{\textbf{Enhancement of the BLS spectra}. \textbf{a},\textbf{b}, BLS spectra acquired on a bare permalloy film (black circles) and on the same film with a 60\,nm thick, 175\,nm wide silicon disk on top (red squares). The black and red solid lines represent fits to the experimental data assuming the Gaussian detection function. The error margins of the experimental data are estimated on the basis of the Poisson distribution. The blue lines show spin wave dispersions for Damon-Eshbach (DE) and backward volume (BV) modes. \textbf{c}, Gaussian detection functions extracted from the fits of the experimental data with and without the nanoresonator in 50\,mT external magnetic field.}
    \label{fig2}
\end{figure}

\section*{Improvement of high-$k$ magnon detection sensitivity}
The dramatic improvement of high-$k$ magnon detection sensitivity in the presence of a 175\,nm wide silicon disk is visible in Figure \ref{fig2}. We can see that at low magnetic field of 50\,mT (Fig. \ref{fig2}a) the BLS signal increases, and the spin wave band broadens towards higher frequencies. At high magnetic field of 550\,mT (Fig. \ref{fig2}b), we can again see the increase of the BLS signal, and the spin wave band now broadens to both sides. 

Different broadening of the spin wave band at high and low magnetic fields can be explained by different shapes of the spin wave dispersion relations of the permalloy thin film for different directions of $k$-vector $\boldsymbol{k}$ with respect to the direction of magnetization vector $\boldsymbol{M}$ \cite{Kalinikos1986, Herring1956}. The upper part of the spin wave band is limited by the Damon-Eshbach mode ($\boldsymbol{k}\perp\boldsymbol{M}$, DE, Figs. \ref{fig2}a,b light blue solid line), which rises at both values of the external field, and for exchange dominated (high-$k$) spin waves converges towards quadratic dependence of frequency $f \propto k^2$. Hence, the shift of the right edge of the detected spin-wave band towards higher frequencies always means an enhanced sensitivity to spin waves with higher $k$-vectors. The left edge of the spin wave band is limited by the backward volume mode ($\boldsymbol{k}\parallel\boldsymbol{M}$, BV, Figs. \ref{fig2}a,b, dark blue dashed line), which first decreases in frequency for dipolar (low-$k$) spin waves and then increases for exchange dominated (high-$k$) spin waves (and again converges towards quadratic dependence $f \propto k^2$). In the low magnetic field, the exchange interaction prevails already for the spin waves with $k \approx 10\,\mathrm{rad}$\,\upmu m$^{-1}$ and the drop in the frequency for the BV mode is only 0.2\,GHz. This results in a sharp increase of the BLS signal at the left edge of the spin wave band. The sharp increase is the same for both measured spectra (with the silicon disk and on the bare film). The \upmu-BLS even without the presence of the dielectric nanoresonator can still detect spin waves with $k$-vectors around the mode minimum at $k \approx 10\,\mathrm{rad}$\,\upmu m$^{-1}$ and thus the complete lower part of the spin wave band is captured in both cases. In the case of high magnetic field, the onset of the exchange dominated spin waves occurs at much higher values of $k$, at approx. $30\,\mathrm{rad}$\,\upmu m$^{-1}$. Here, the BV mode is very pronounced, and the mode frequency decreases approx. 1.5\,GHz down from the ferromagnetic resonance frequency (FMR, $k = 0\,\mathrm{rad}$\,\upmu m$^{-1}$) before it starts rising again (Fig. \ref{fig2}b, dark blue dashed line). In this case, the \upmu-BLS on the bare film cannot detect spin waves above $k = 10\,\mathrm{rad}$\,\upmu m$^{-1}$ and capture the whole lower part of the spin wave band.

To quantify the enhancement of the BLS signal, we use a simple phenomenological model which assumes a Gaussian detection function $\Gamma(k)$ in the $k$-space (see Methods)
\begin{equation}
\Gamma= A \exp\left(\frac{-k^2}{2\left(\mathrm{HWTM}/\sqrt{2\ln{10}}\right)^2}\right) + \mathrm{bg}.    
\end{equation}

The detection function has two parameters which we use for the quantification of the enhancement. Amplitude $A$, which represents the strength of the BLS signal, and half-width-at-tenth-maximum $\mathrm{HWTM}$, which represents the maximum detectable $k$-vector. The third parameter $\mathrm{bg}$ representing the background signal is not used in further analysis. By multiplying this detection function by the spin-wave density of states in the magnetic thin film $\mathcal{D}(f, k_x, k_y)$, obtained from a micromagnetic simulation, we can model the acquired BLS signal $\sigma_{\mathrm{BLS}}\left(f\right)$ 

\begin{equation}
\sigma_\mathrm{BLS}\left(f\right)=\iint_{k_x,k_y}{\mathcal{D}\left(f,k_x,k_y\right)\Gamma\left(k_x,k_y\right)\mathrm{d}k_x\mathrm{d}k_y}.
\label{eq:PhenoModel}
\end{equation}

By fitting the parameters $A$ and $\mathrm{HWTM}$, we can obtain very good agreement between the model and the experimentally measured spectra for both the bare film and the silicon-disk-enhanced measurement (see black and red solid lines in Fig. \ref{fig2}a,b). In Fig. \ref{fig2}c, the detection function $\Gamma(k_x)$ resulting from the fit to the experimental data measured  at 50\,mT is plotted for the bare film (black line) and for the measurement on the silicon disk (red line). This figure gives us a direct visualization of the enhancement of the detection sensitivity caused by the presence of the dielectric nanoresonator. The $\mathrm{HWTM}$ (i.e. the maximum detectable $k$) increased from $9.5 \pm 1.0$\,$\mathrm{rad}$\,\upmu m$^{-1}$ for bare film to $47 \pm 3$\,$\mathrm{rad}$\,\upmu m$^{-1}$ for silicon disk, whereas the $A$ increased from $245 \pm 6$\,cts to $259 \pm 19$\,cts. Note that the increase in the amplitude of the Gaussian function does not represent the total increase of the integrated signal. If we integrate the detection function, we obtain the value of $2720 \pm 170$\,$\mathrm{cts}\,\mathrm{rad}$\,\upmu m$^{-1}$ for the case of the bare film and the value of $14300 \pm 1600\,\mathrm{cts}\,\mathrm{rad}$\,\upmu m$^{-1}$ for the silicon-disk-enhanced signal. This gives us an enhancement factor of $5.3$.  

\begin{figure*}
    \centering
    \includegraphics{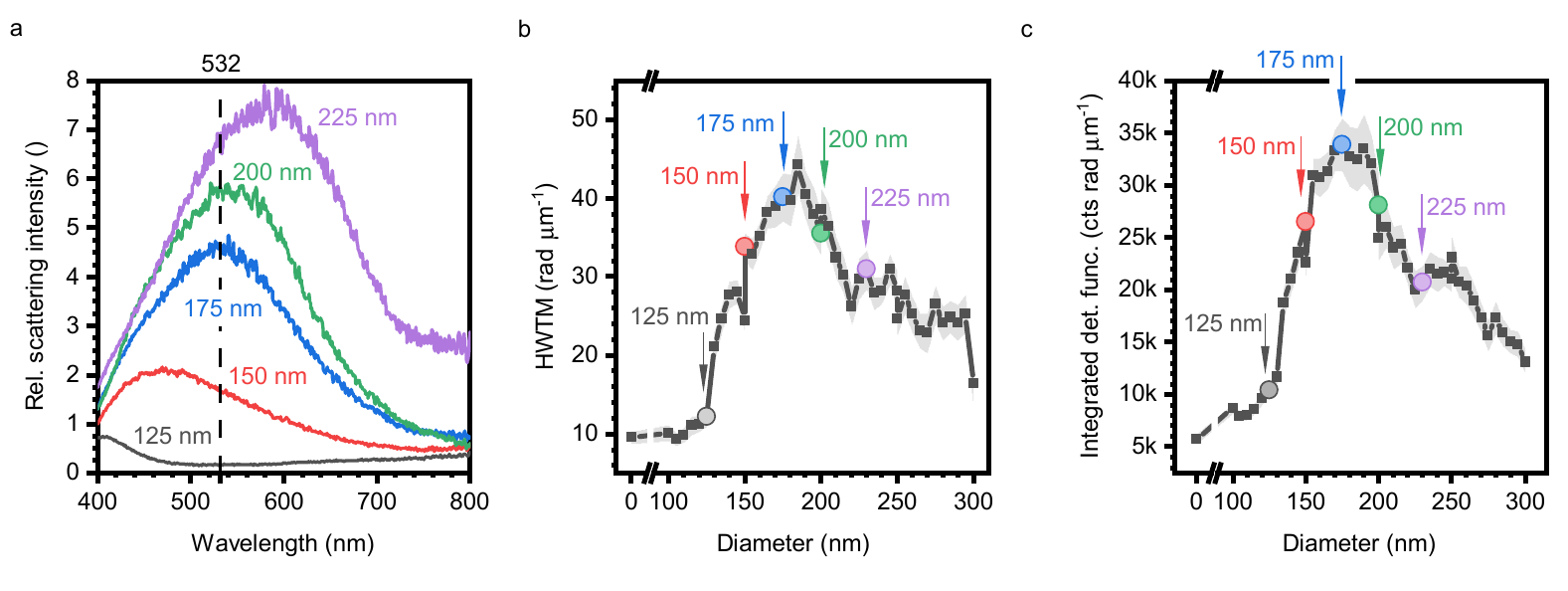}
    \caption{\textbf{Dependence of the BLS enhancement on the silicon disk diameter.}  \textbf{a}, Relative scattering intensities of five selected silicon disks measured using dark-field optical spectroscopy. The dashed line marks the 532\,nm wavelength of the \upmu-BLS laser. \textbf{b, c,} Dependence of HWTM (half-width-at-tenth-maximum, maximum detectable $k$-vector) and integrated detection function (BLS signal enhancement) on the diameter of the silicon disk. The arrows and coloured points refer to the diameters with elastic scattering spectra shown in panel \textbf{a}.}
    \label{fig3}
\end{figure*}

Mie resonances are strongly dependent on the geometry of nanoresonators. To investigate this dependency, we have measured a series of silicon disks with diameters ranging from 100 to 300\,nm. Fig. \ref{fig3}a shows relative scattering intensities acquired by dark-field optical spectroscopy. We can see a characteristic red shift of the Mie resonances with increasing disk diameter. For 175\,nm disk the peak resonance wavelength perfectly matches with the laser in our \upmu-BLS setup (532\,nm). To quantify the dependence of the enhancement of the measured BLS spectra on the disk diameter, we fitted the HWTM (Fig. \ref{fig3}b) and $A$ (Fig. \ref{fig3}c) parameters for each disk diameter. From these data, we can see that the enhancement of both parameters starts appearing for disk diameters beyond approx. 125\,nm and reaches its maximum for the diameters between 170 and 200\,nm. When the disk diameter exceeds 200\,nm, we observe a sharp decrease of both parameters.

\section*{Theoretical description of Mie-resonance enhanced BLS}
To further understand the mechanism of the enhancement, we implemented the so-called continuum model of inelastic scattering \cite{Landau1960}. In this model, the driving electric field $\boldsymbol{E}_\mathrm{d}$ probes the dynamic modulation of the susceptibility via magneto-optical coupling, which gives the polarization inside the magnetic material as \cite{Cottam1986, Cottam1976, Qiu2000}
\begin{equation}
\boldsymbol{P}(t,\boldsymbol{r})=\hat{\boldsymbol{\chi}} (t,\boldsymbol{r}) \boldsymbol{E}_\mathrm{d} (t,\boldsymbol{r}),
\end{equation}
where $\hat{\boldsymbol{\chi}}=\hat{\boldsymbol{\chi}}_\mathrm{mat} + \hat{\boldsymbol{\chi}}_\mathrm{SW}(t,\boldsymbol{r})$ is a sum of the static material susceptibility $\hat{\boldsymbol{\chi}}_\mathrm{mat}$ and of the additional dynamic contribution caused by spin waves $\hat{\boldsymbol{\chi}}_\mathrm{SW}(t,\boldsymbol{r})$. It should be stressed that the modulations in the susceptibility caused by magnons are on a vastly different time scale than the optical cycle of the probing photons.
As a consequence, there is mixing of frequencies both in temporal and spatial domains, namely

\begin{equation}
\label{Pqvec}
\boldsymbol{P}(\omega,\boldsymbol{k_\mathrm{p}},z)= \hat{\boldsymbol{\chi}} (\omega_{\mathrm{m} },\boldsymbol{k}_{\mathrm{m} },z) \boldsymbol{E}_\mathrm{d} (\omega - \omega_{\mathrm{m} },\boldsymbol{{k_\mathrm{p}}} - \boldsymbol{k_\mathrm{m}} ,z) ,
\end{equation}

\noindent
where $\omega$ denotes the frequency of the induced polarization, $\boldsymbol{k_\mathrm{p}}$ stands for its in-plane wavevector (parallel to the permalloy layer), while $\omega_{\mathrm{m} }$ and $\boldsymbol{k}_{\mathrm{m} }$ represent their magnon counterparts. 

The polarization vector in~(\ref{Pqvec}) acts as a local source of radiation that eventually forms the detected BLS signal. The contribution of a particular spatial frequency to this signal is determined by its ability to efficiently couple to the free space continuum and pass through the optical setup towards the detector. In the case of a bare permalloy layer without any particles, the range of spatial frequencies that can reach the detector is mostly limited by the numerical aperture of the used objective. Presence of a silicon disk (or any other perturbation) broadens this frequency space by providing an extra momentum that is required for high-$k_\mathrm{p}$ fields to be scattered into the free space continuum. This scattering process thus represents
an additional channel by which the information from local sources can reach the detector.

The transition from near-field to far-field can be mathematically expressed using Green's function formalism

\begin{equation}
\label{Dyad}
\boldsymbol{E}_{\mathrm{FF} }(\omega,\boldsymbol{k_\mathrm{p}})= \hat{\boldsymbol{G}} (\omega,\boldsymbol{k_\mathrm{p}},\boldsymbol{k_\mathrm{p}}^{\prime}) \boldsymbol{P}(\omega,\boldsymbol{k_\mathrm{p}}^{\prime}) .
\end{equation}

\noindent
The dyadic Green's function $\hat{\boldsymbol{G}} (\omega,\boldsymbol{k_\mathrm{p}},\boldsymbol{k_\mathrm{p}}^{\prime})$ embodies the response of a system to a local source and it accounts fully for the presence of any scattering object or substrate effects. 
By inserting (\ref{Pqvec})~into~(\ref{Dyad}) and integrating over all spatial frequencies supplied by the driving electric field, the far-field angular spectrum becomes

\begin{equation}
\begin{aligned}
\label{Farfield}
\boldsymbol{E}_{\mathrm{FF} }(\omega_{\mathrm{m} } &,\boldsymbol{k_\mathrm{m}}, \omega,\boldsymbol{k_\mathrm{p}})=  \int \mathrm{d}^2 k_\mathrm{p}^{\prime} \,\, \hat{\boldsymbol{G}} (\omega,\boldsymbol{k_\mathrm{p}},\boldsymbol{k_\mathrm{p}}^{\prime})  \\
& \hat{\boldsymbol{\chi}} (\omega_{\mathrm{m} },\boldsymbol{k}_{\mathrm{m} }) \boldsymbol{E}_\mathrm{d} (\omega, \boldsymbol{k_\mathrm{p}}^{\prime} - \boldsymbol{k}_{\mathrm{m} }).
\end{aligned}
\end{equation}

Another important aspect of the BLS detection process is the limited area from which the signal is collected 
Assuming that the collection spot has a Gaussian spatial profile $h(x,y) = e^{- (x^2+y^2)/w_{\mathrm{c}}^{2} }$, the detectable portion of the far-field radiation amounts to

\begin{equation}
\label{Farfield2}
\boldsymbol{E}_{\mathrm{FF} }(\boldsymbol{r}_{\parallel})= h(\boldsymbol{r}_{\parallel}) \!\!\!\!\!\! \int\limits_{k_\mathrm{p} \leq k_{0} \mathrm{NA} } \!\!\!\!\!\! \mathrm{d}^2 k_\mathrm{p} \,\, e^{\mathrm{i} \boldsymbol{k_\mathrm{p}} \cdot \boldsymbol{r}_{\parallel}} \, \boldsymbol{E}_{\mathrm{FF} }(\boldsymbol{k_\mathrm{p}}),
\end{equation}

\noindent
where the integration limits reflect the restrictions placed on the spatial frequencies by the numerical aperture of the objective lens.

Finally, to estimate the strength of the BLS signal at a particular frequency $\omega_{\mathrm{m} }$, one has to sum up contributions from all magnons (i.e. integrate over $\boldsymbol{k}_{\mathrm{m} }$). 
Hence, the modelled BLS signal reads

\begin{equation}
\begin{aligned}
\label{sigma}
\sigma & (\omega_{\mathrm{m} } )  = \! \int \!\! \mathrm{d}^2 r_{\parallel} \int \!\! \mathrm{d}^2 k_{\mathrm{m} } \bigg\vert \,\, h(\boldsymbol{r}_{\parallel}) \!\!\!\!\!\!\! \int\limits_{k_\mathrm{p} \leq k_{0} \mathrm{NA} } \!\!\!\!\!\!\! \mathrm{d}^2 k_\mathrm{p} \, e^{\mathrm{i} \boldsymbol{k_\mathrm{p}} \cdot \boldsymbol{r}_{\parallel}}  \\
& \int \! \mathrm{d}^2 k_\mathrm{p}^{\prime} \,\, \hat{\boldsymbol{G}} (\boldsymbol{k_\mathrm{p}},\boldsymbol{k_\mathrm{p}}^{\prime}) \, \hat{\boldsymbol{\chi}} (\omega_{\mathrm{m} },\boldsymbol{k}_{\mathrm{m} }) \boldsymbol{E}_\mathrm{d} (\boldsymbol{k_\mathrm{p}}^{\prime} - \boldsymbol{k}_{\mathrm{m} }) \bigg\vert^{2}.
\end{aligned}
\end{equation}

\noindent
Inspecting the above expression, there are apparently many factors that can affect the resulting shape of the measured BLS spectrum, but the ability to access high-$k$ magnons has only two possible origins: scattering of high-$k$ field components into the free space continuum represented by $\hat{\boldsymbol{G}} (\boldsymbol{k_\mathrm{p}},\boldsymbol{k_\mathrm{p}}^{\prime})$ or the fact that silicon disks broaden the range of spatial frequencies making up the driving electric field $\boldsymbol{E}_\mathrm{d} (\boldsymbol{k_\mathrm{p}})$. While one of these processes occurs before and the other one after the magnon scattering event, both of them are facilitated by the silicon disks and their near-fields.

\begin{figure*}
    \centering
    \includegraphics{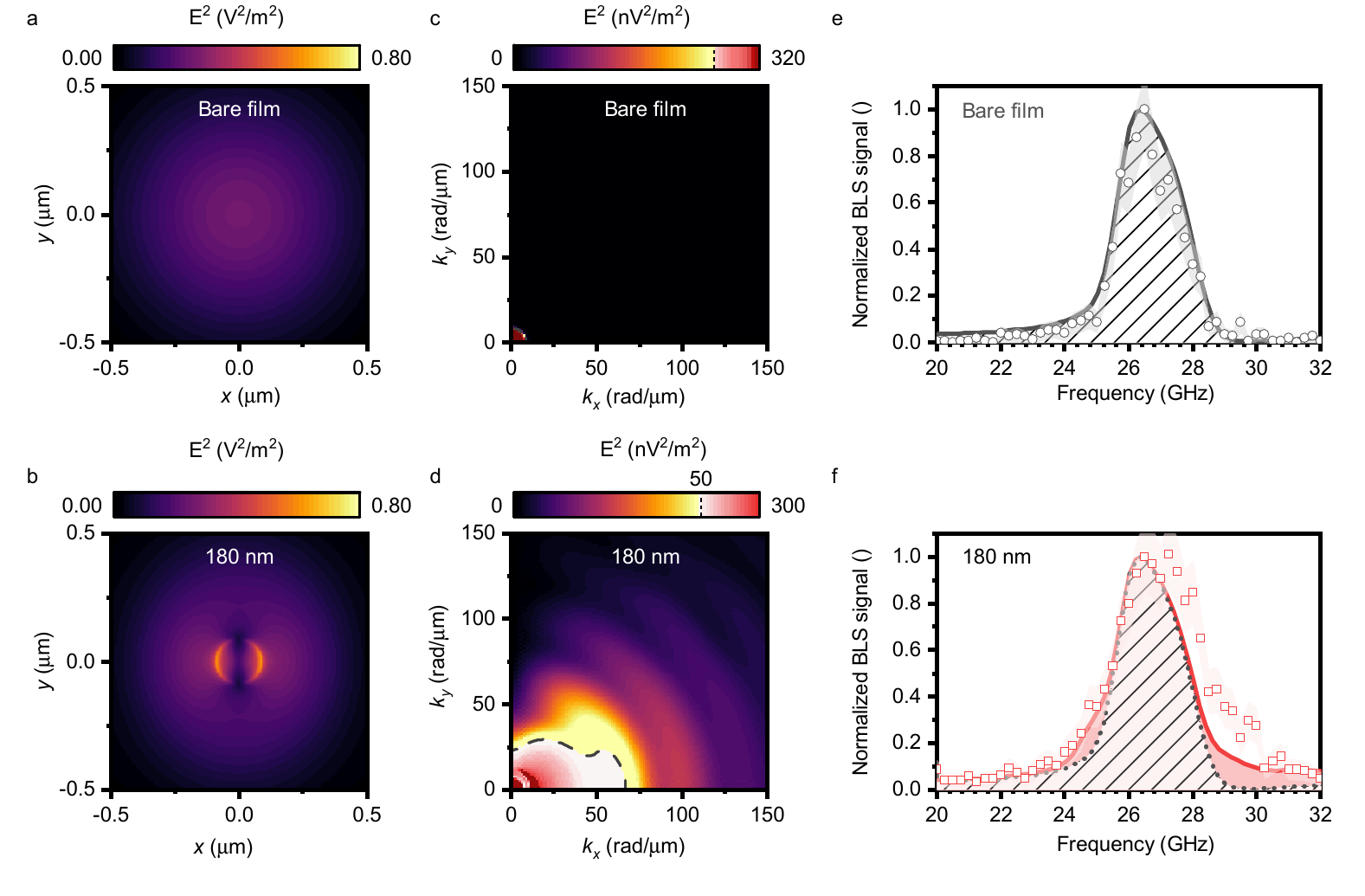}
    \caption{\textbf{Numerical simulation of the electric field distribution and theoretically calculated BLS signal.} \textbf{a, b} Real-space distribution of squared electric field  for \textbf{(a)}  bare permalloy film and \textbf{(b)} for the same permalloy film with 180\,nm  wide silicon disk on top. \textbf{c, d} Reciprocal space distributions of the data shown in \textbf{(a), (b)}. The colormap has two linear regions; 0--50 V$^2$m$^{-2}$ and 50--320 V$^2$m$^{-2}$ for better clarity. The boundary between the regions is marked by the dashed line.  \textbf{e, f} Calculated BLS signal compared to the experimental data for \textbf{(e)} bare permalloy film and for the same permalloy film with \textbf{(f)} 180\,nm wide silicon disk. The calculated BLS signal is normalized to the highest value.}
    \label{fig4}
\end{figure*}

To obtain the distribution of the driving electric field inside the sample, we employed finite-difference-time-domain (FDTD) simulations. We used Gaussian illumination with experimentally obtained beam parameters (see Extended Data Fig. \ref{Extfig-beamShape}). In the bare film, the electric field has a 2D Gaussian distribution (Fig. \ref{fig4}a), and its Fourier transform is shown in Fig. \ref{fig4}c. In the case silicon disk, we observe the field localized in subdiffraction regions due to Mie resonance (see Fig. \ref{fig4}b). Consequently, the electric field spans over a larger area of the $k$-space  exceeding values of 120\,$\mathrm{rad}$\,\upmu m$^{-1}$ (see Fig. \ref{fig4}d).

This model allows us to directly calculate the theoretical BLS signal (see Figs. \ref{fig4}e-f). In the case of bare film, the model perfectly matches to the experimental spectra.
Figure \ref{fig4}b shows that silicon disk supports dipole Mie resonance (see also Extended Data Fig. \ref{Extfig-180vs220} for disk cross sections). In reciprocal space (Fig. \ref{fig4}d) we can see how the electric field localization provides an extra momentum that is required for high-$k$ magnon detection. The modelled BLS spectra (\ref{fig4}f) is in a good agreement with the experimental data until the frequency reaches approx. 30\,GHz (which corresponds to $k$=30\,$\mathrm{rad}$\,\upmu m$^{-1}$), above this value the model overestimates the BLS signal strength. The lower sensitivity to higher $k$-vectors in real experiments can be caused e.g. by unstable laser spot position and imperfect polarization of the incoming laser beam.

\begin{figure*}
    \centering
    \includegraphics{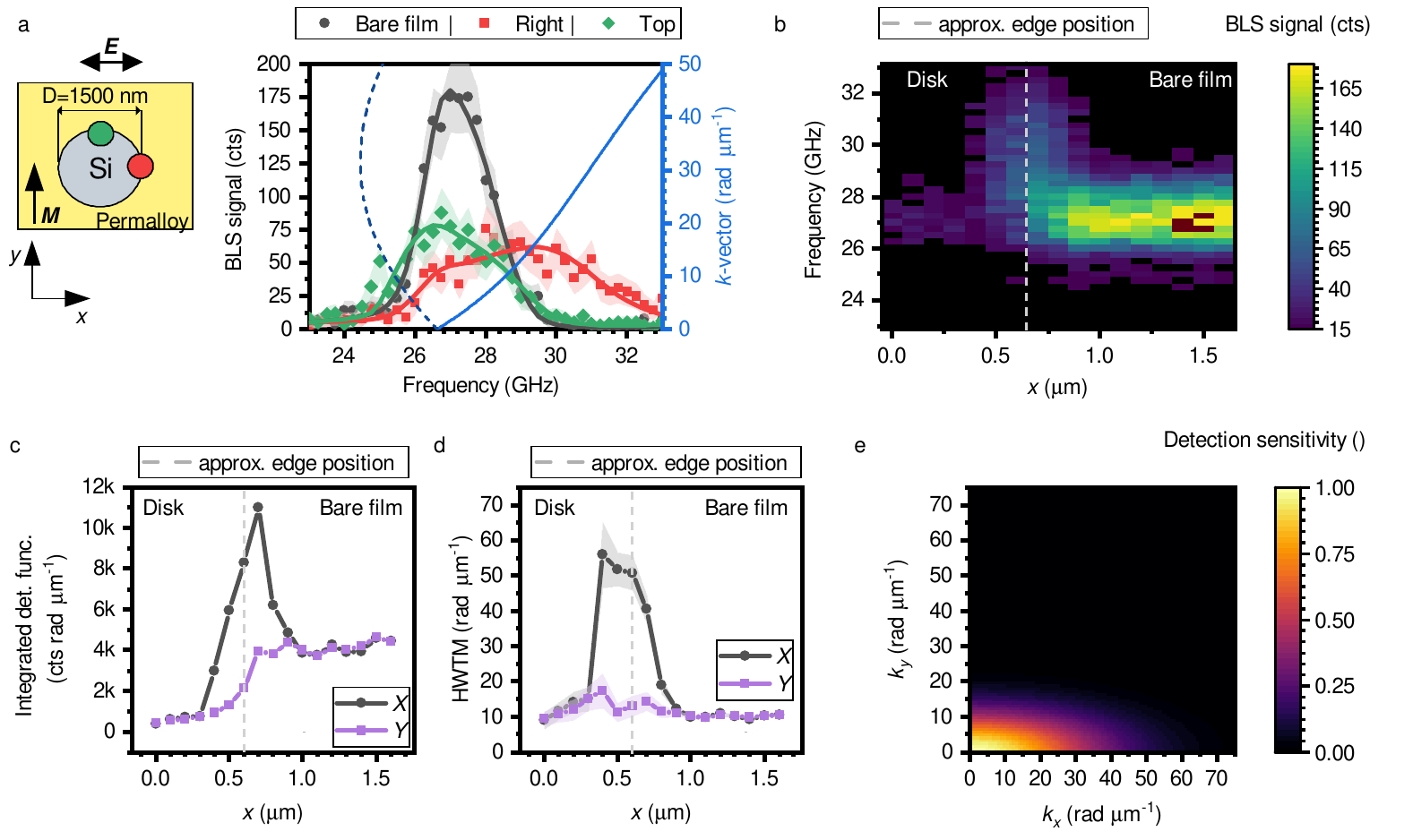}
    \caption{\textbf{BLS enhancement at the edges of a large silicon disk.} \textbf{a,} BLS spectra acquired in the field of 550\,mT on the right edge (red line), and on the top edge (green line) of a large silicon disk compared with the spectra measured on a bare permalloy film (black line) with magnetization $M$ pointing in the $y-$direction, see schematic. Blue lines represent analytically calculated dispersion relation for DE (light blue solid line) and BV (dark blue dashed line) spin waves. The error band is calculated from the Poisson distribution.  \textbf{b,}  BLS line scan across the right edge of the large silicon disk. The edge position is marked by the light gray dashed line. \textbf{c, d,} Integrated detection function and HWTM of the detection function extracted from a fit to the data presented in \textbf{(b)}. \textbf{ e,} 2D detection function extracted from a fit to the data measured at the right edge of the disk.}
    \label{fig5}
\end{figure*}

In resonant photonic structures, the areas with locally enhanced electromagnetic field are often called hot spots as the temperature in their vicinity can rise by hundreds of Kelvins upon illumination. Such large increase in the temperature could change the magnetic properties of the studied material. This behavior is typical for plasmonic structures made of metals. To investigate how much can the silicon disk heat up the neighboring magnetic layer, we iteratively solved heat transfer and Maxwell equations. This calculation reveals, that for the used incident laser power of 3\,mW, the temperature increase in the hot spots is max. 30\,K (see Extended Data Fig. \ref{Extfig-Temp}). This reduces the saturation magnetization of the permalloy by 0.6\,percent, which in turn shifts down the spin wave band by approx. 20\,MHz, which is far below the resolution of our TFPi with the mirror spacing set to 3\,mm \cite{Lindsay1981}.

\section*{Directional sensitivity}
To estimate how the laser spot positioning influences the measurement, we performed 2D mapping of (1$\times$1)\,\upmu m$^2$ area around the nanoresonator, see Extended Data Fig. \ref{Extfig-2Dscan}. To achieve the maximum enhancement, the laser spot has to be aligned with sub-100\,nm precision on the centre of the nanoresonator.

To explore the role of the nanoresonators' edges, we have measured BLS spectra with the laser spot focused on the right edge and then on the top edge of a large, 1500\,nm wide, disk (see Fig. \ref{fig5}a). We can see that compared to the measurement on a bare film, the overall BLS signal is lower at both laser spot positions. Nevertheless, the enhancement of maximum detectable $k$-vector is still present. Interestingly, when the laser beam is positioned on the right edge of the disk we can see broadening of the spin wave band towards higher frequencies (Fig. \ref{fig5}a, red line). When the laser beam is positioned on the top edge of the cylinder we can see broadening of the spin wave band towards lower frequencies (Fig. \ref{fig5}a, green line). This experiment suggests that it is possible to change the sensitivity to different spin wave $k$-vector directions. On the right edge, we were more sensitive to DE spin waves ($\boldsymbol{k}\perp\boldsymbol{M}$), whereas on the top edge, we were more sensitive to BV spin waves ($\boldsymbol{k}\parallel\boldsymbol{M}$).  

To quantify this directional sensitivity, we performed a BLS line scan from the disk center across the right edge of the disk (Fig. \ref{fig5}b). Then we fitted the measured data with Eq. \ref{eq:PhenoModel}, considering different HWTM parameters for $k_x$ and $k_y$. The fitted detection function in Fig. \ref{fig5}e indeed confirms that the HWTM parameter and integrated detection function is enhanced only in the vicinity of the disk edge and only in the $x$-direction (Fig. \ref{fig5}c,d).

\section*{Detection of coherently excited spin waves}
\begin{figure*}
    \centering
    \includegraphics{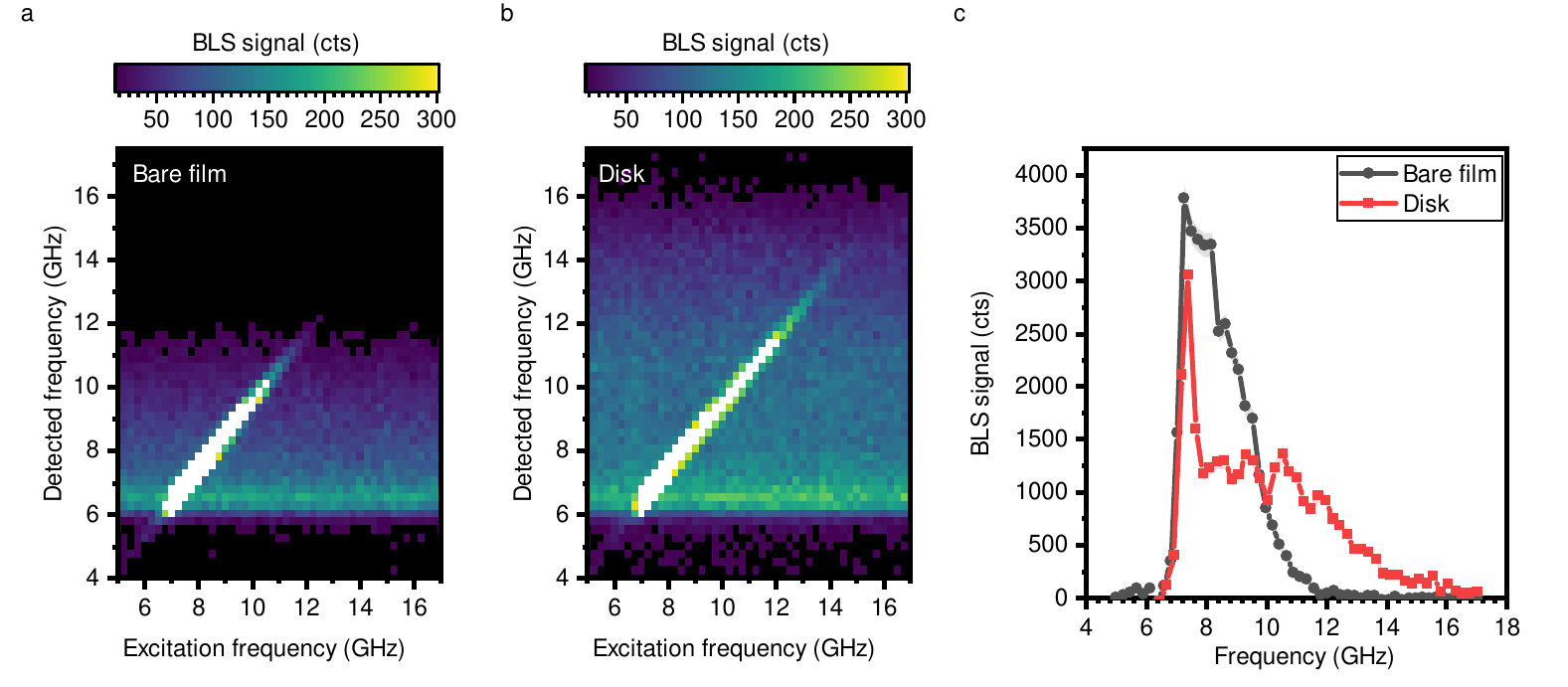}
    \caption{\textbf{Enhancement of BLS signal from coherently excited spin waves.} \textbf{a, b,} Spin wave spectra acquired $1\,$\upmu m  away from the excitation antenna, measured on a bare film \textbf{(a)}, and on a 180\,nm wide silicon disk \textbf{(b)}. Note, that the BLS signal from thermal spin waves is visible for all excitation frequencies, whereas coherently excited spin waves manifest themselves as a strong signal on the diagonal. \textbf{c,} Signal from coherent spin waves extracted from the diagonal in \textbf{(a), (b)} after subtraction of the thermal background.}
    \label{fig6}
\end{figure*}

So far, we have investigated only thermal spin waves, but for many experiments in magnonics one needs to probe coherently excited spin waves with a defined $k$-vector. To prove that our approach works also for coherent spin waves, we fabricated a sample with 180\,nm wide microwave (RF) antenna in the vicinity of 200\,nm wide silicon disks on the top of the permalloy thin film (see Extended Data Fig. \ref{Extfig-SEM}). We connected the antenna to a RF generator with the excitation power set to 10\,dBm, and swept the excitation frequency from 5\,GHz to 17\,GHz. The external magnetic field for this experiment was set to 50\,mT and we probed the spin waves at the distance 1\,\upmu m from the antenna. In the case of the bare film (Fig. \ref{fig6}a), the excitation efficiency of the antenna is broader than the detection sensitivity, which is confirmed by the fact that the signal from coherent spin waves ends at the same frequency as the thermal background. Contrary, in the case of the silicon disk (Fig. \ref{fig6}b), the thermal background exceeds the coherent spin waves, which means that the coherent spin waves were limited by the excitation efficiency of the antenna. For the antenna with the width $w=180$\,nm the frequency cut-off is approximately $w/2\pi$ which corresponds to the $k$-vector of 29\,$\mathrm{rad}$\,\upmu m$^{-1}$ \cite{Vavnatka2021, Vlaminck2010}. This is consistent with the maximum detected frequency of 16\,GHz, corresponding to $k= 30\,\mathrm{rad}$\,\upmu m$^{-1}$.

\section*{Conclusion}
Our results show that with the use of dielectric nanoresonators it is possible to use the \upmu-BLS technique to detect thermal and coherently excited magnons with $k$-vectors exceeding $50\,\mathrm{rad}$\,\upmu m$^{-1}$ (which corresponds to spin wave wavelength $\lambda=125$\,nm). We assert that achieving detection of spin waves with even higher $k$-vectors is just a signal-to-noise issue, as the simulated Fourier images of nanoresonator-enhanced electric field distributions show nonzero intensity even for $k > 120\,\mathrm{rad}$\,\upmu m$^{-1}$ (see Fig. \ref{fig4}). Spin waves with $k$-vectors higher than $50\,\mathrm{rad}$\,\upmu m$^{-1}$ could be measured using a source of coherent spin waves with better excitation efficiency at high $k$-vectors \cite{Chernov2020, Chen2021, Liu2018, Che2020, Hamalainen2017} as this should substantially improve the signal-to-noise ratio. Also, the geometry and the material of the dielectric nanoresonator can be further optimized in order to excite higher order Mie or anapole resonances. This should lead to further increase of the enhancement factor \cite{Kuznetsov2012, Haar2016, Yang2018}. These findings elevate \upmu-BLS to the forefront of the nanoscale magnonics research and the possibility to probe materials with high momentum photons is relevant also for other applications, e.g. for phononic studies \cite{smith2021} or even mechanobiology experiments \cite{prevedel2019}.

\backmatter

\bmhead{Acknowledgments}
CzechNanoLab project LM2018110 is gratefully acknowledged for the financial support of the measurements and sample fabrication at CEITEC Nano Research Infrastructure. O.W. was supported by Brno PhD talent scholarship. O.W. and J.H. acknowledge support from the project Quality Internal Grants of BUT (KInG BUT), Reg. No. CZ.02.2.69 / 0.0 / 0.0 / 19\_073 / 0016948, which is financed from the OP RDE. F.L. acknowledges the support by the Grant Agency of the Czech Republic (21-29468S). We acknowledge R. Dao and Nenovison company for help with AFM measurements.


\bmhead{Availability of data and materials}

The datasets generated during and/or analyzed during this study are available from the corresponding authors upon reasonable request.

\bmhead{Code availability} 

The Matlab code used to evaluate the BLS enhancement and to model the BLS spectra using the input from FDTD simulations is available on Github together with COMSOL mph file \cite{github}. The code used for the dispersion relation calculation is available on Github in a separate repository \cite{githubSWT}.

\bmhead{Authors' contributions}

O.W. and M.U. designed the experiments, performed and evaluated the BLS measurements. M.H., O.W., and T.S. developed the theoretical scattering model. J.K., M.D., K.D., and M.S. prepared the samples and characterized them by SEM and AFM. F.L. and T.S. performed scattering measurements. J.H. and O.W. performed VSM and FMR measurements. J.Z. performed COMSOL multiphysics simulations. All authors contributed in writing and reviewing the paper.


%
%
%
%

\bibliography{sn-bibliography}
\begin{appendices}
\newpage
\section*{Methods}
\textbf{Sample fabrication}
Samples with silicon disks were fabricated by electron beam lithography and a lift-off process. We started with room temperature deposition of a 30\,nm thick permalloy film onto a Si(100) substrate using e-beam evaporation from Ni$_{80}$Fe$_{20}$ (at. \%) pellets. Then we spin coated a double-layer polymethyl methacrylate resist (200\,nm thick Allresist AR-P 649.04 200K and 60\,nm thick AR-P 679.02 950K, this resist combination provides sufficient undercut for lift-off). The patterns were written with RAITH 150-two and Tescan Mira e-beam writers. Silicon film with the thickness of 60\,nm was consequently deposited onto the patterned sample by RF magnetron sputtering from a crystalline silicon target at room temperature. The lift-off procedure consisted of immersing the sample in acetone for 8.5\,h, followed by 30 second isopropanol rinse and blow-drying by nitrogen gas. The antennas for coherent excitation were prepared in a second lithographic step, again using the double-layer resist and the lift-off process. The antennas consisted of a (10\,nm SiO$_2$/85\,nm Cu/10\,nm Au) multilayer stack deposited all at once in the e-beam evaporator.

The sample with silver nanospheres was prepared by drop-casting from a commercial solution (nanoComposix) to a 30\,nm thick permalloy layer covered by 2\,nm of an insulating Al$_2$O$_3$ spacer prepared by atomic layer deposition. We deposited $30\,$\upmu$\mathrm{l}$ droplet of the solution containing silver nanospheres with the diameter of 200\,nm to the sample surface and let the sample dry for 120\,minutes. After that, the solution was rinsed in deionized water and blow-dried by clean dry air.

After the fabrication, all samples were checked for their exact shape, size and uniformity by scanning electron microscopy (Tescan Lyra) and atomic force microscopy (Bruker Dimension Icon). The scanning electron microscope image of the sample with 175\,nm wide silicon disks and 200\,nm wide RF excitation antenna is shown in Extended data Fig. \ref{Extfig-SEM}.

\textbf{BLS experiments}
We used Cobolt Samba solid state laser with the wavelength of 532\,nm and maximal optical power of 300\,mW. The laser output pinhole was protected against the back-reflections with Faraday optical isolator. Spectral purity of the laser light was improved by a Fabry-Perot filter (TCF-2, TableStable). After the filter, a small portion of the light was guided towards the TFPi to ensure its stabilization during the measurement. The rest of the light passed through a $\lambda/2$ waveplate and a polarizer. The $\lambda/2$ waveplate was mounted on a motorized stage and served for setting the incident laser beam power. Unless stated otherwise, we always set the optical power to the value of 3\,mW. We used an optical microscope with active stabilization for compensation of the mechanical drifts of the sample (THATec Innovation). The incident laser light was guided through two 50:50 beam splitters towards Zeiss LD EC Epiplan-Neofluar 100$\times$/0.75 BD objective lens. The scattered light (including the portion which underwent the BLS process) was collected by the same objective lens and guided through the same 50:50 beam-splitters back towards other $\lambda/2$ waveplate and then into the TFPi. We used TFP-2HC interferometer (Table Stable) \cite{Lindsay1981} with the resolution of approx. 250\,MHz with mirror spacing set to 3\,mm. We used 450\,\upmu m input and 700\,\upmu m output pinholes. For the thermal spectra we acquired 2000\,scans, while in the case of coherent excitation we measured 2928\,scans. In both cases, the acquisition time was 1\,ms for a single frequency bin. The acquisition of one thermal spectrum took approx. 10\,minutes.
To generate the magnetic field, we used a water-cooled electromagnet GMW\,5403 driven by two current sources KEPCO BOP20-20DL. The magnetic field in the sample position was measured by a LakeShore 450 Hall probe. 

\textbf{Quantification of the BLS signal enhancement}
To quantify the enhancement of the BLS signal, we developed a simple phenomenological model based on the following equation
\begin{equation*}
\sigma_{BLS}\left(f\right)=\iint_{k_x,k_y}{\mathcal{D}\left(f,k_x,k_y\right)\Gamma\left(k_x,k_y\right)dk_xdk_y},    
\label{eq-phen-mod}
\end{equation*}
where $\mathcal{D}\left(f,k_x,k_y\right)$ is the density of states of spin waves, $\Gamma\left(k_x,k_y\right)$ is an instrumental detection function and $\sigma_{BLS}\left(f\right)$ is the measured signal. We assume that the detection function $\Gamma$ has a Gaussian form 
\begin{equation*}
\begin{aligned}
\Gamma={} & A \exp\left(\frac{-k_x^2}{2\left(\mathrm{HWTM}_x/\sqrt{2\ln{10}}\right)^2}\right)  \\
& \exp\left(\frac{-k_y^2}{2\left(\mathrm{HWTM}_y/\sqrt{2\ln{10}}\right)^2}\right) +\mathrm{bg},
\end{aligned}
\end{equation*}
where $A$ is the strength of the measured signal, $k_x$ $(k_y)$ is the spin-wave $k$-vector in $x$ $(y)$ direction, and $\mathrm{HWTM}_x\ \left(\mathrm{HWTM}_y\right)$ is half width at tenth of maximum of the detection sensitivity for spin-wave $k$-vectors in $x\left(y\right)$ direction. The bg stands for the background signal, which could be caused by a dark current in the detector, inelastic scattering on the phonon modes or by stray light. To quantify the enhancement of the BLS signal in our data, we only fit $A$ and HWTM.

The density of states $\mathcal{D}\left(f,k_x,k_y\right)$ was obtained from a micromagnetic simulation, using MuMax3 micromagnetic solver \cite{Vansteenkiste2014}. The simulation size was set to $7 680\times7 680\times34.8$\,nm$^3$ with cell size of 3.75\,nm in in-plane and 4.35\,nm in out-of-plane direction. Periodic boundary conditions with 32 repetitions in both in-plane directions were used. Saturation magnetization ($M_\mathrm{s}$\,=\,741\,kA\,m$^{-1}$), gyromagnetic ratio ($\gamma$\,=\,29.5\,GHz\,T$^{-1}$), and exchange constant ($A_\mathrm{ex}$\,=\,16\,pJ\,m$^{-1}$) used in the simulation were extracted from a ferromagnetic resonance measurements (see Extended Data Fig. \ref{Extfig-FMR}). The initial magnetization $\boldsymbol{m}$ and the external field pointed in the $x$ direction. The dynamics of the magnetization was excited by a 3D sinc pulse $B_\mathrm{sinc}\left(x,y,t\right)$, which is depicted in Extended Data Fig. \ref{Extfig-SimMuMax}a, b. The cut-off frequency was set to 60\,GHz and the cut-off $k$-vector to 150\,$\mathrm{rad}$\,\upmu m$^{-1}$. The amplitude of the sinc function was 1\,mT, and the pulse was delayed by 100\,ps.

We let the magnetization evolve for 5\,ns with the sampling interval of 8\,ps. The obtained $m_z$ is shown in  Extended Data Fig. \ref{Extfig-SimMuMax}c, d. All three magnetization vector components were taken from the top-most layer, so we acquired a 3D array  $\boldsymbol{m}\left(x,y,t\right)$. The used script code is available at \cite{github}. The $m_z$ component was then transformed to the reciprocal space using a built-in FFT function in Matlab2021a. The obtained dispersion relation was compared with analytical calculation using Kalinikos-Slavin formula \cite{Kalinikos1986, githubSWT}, as it is shown in Extended Data Fig. \ref{Extfig-SimMuMax}e, f. No windowing nor detrending was used. The resulting density of states was then obtained as
\begin{equation*}
\mathcal{D}\left(f,k_x,k_y\right)=n\left(f\right)\ m_z\left(f,k_x,k_y\right)^2,    
\end{equation*}
where $n\left(f\right)$ is Bose-Einstein distribution, which can be calculated as
\begin{equation*}
n\left(f\right)=\ \ \frac{1}{\exp{\left(\frac{2\pi\hbar f}{k_BT}\right)}-1}
\end{equation*}
This distribution at room temperature is shown in  Extended Data Fig. \ref{Extfig-SimMuMax}g. Following Eq. \ref{eq-phen-mod} the detection function ($\Gamma$) with free parameters $A$, HWTM$_x$, and HWTM$_y$ was multiplied by the density of states ($\mathcal{D}$), integrated over $k$-space and fitted to the experimental data using Non-linear Fitting Tool in Matlab2021a. The used function is available at  \cite{github}. For all experiments, where the beam spot was positioned in the center of the silicon disk, we set $k_x=k_y$ and fitted only one HWTM parameter.

\textbf{Analytical modelling of the BLS spectra}

To analytically model the BLS signal generated by a system containing a scatterer (e.g. a silicon disk), one requires several key ingredients that need to be combined and evaluated using Eq.~(\ref{sigma}): the distribution of the driving electric field $\boldsymbol{E}_\mathrm{d} (\boldsymbol{r})$, the susceptibility tensor $\hat{\boldsymbol{\chi}} (\omega_{\mathrm{m} },\boldsymbol{k}_{\mathrm{m} })$, and the dyadic Green's function $\hat{\boldsymbol{G}} (\boldsymbol{k_\mathrm{p}},\boldsymbol{k_\mathrm{p}}^{\prime})$. The complexity of the evaluation itself depends on a number of parameters, namely the numerical aperture of the objective lens ($\mathrm{NA}=0.75$), the size of the collection spot ($w_{\mathrm{c}}=440$\,nm), and the size of the reciprocal space involved in the integrations over $\boldsymbol{k_\mathrm{p}}^{\prime}$ and $\boldsymbol{k_\mathrm{m}}$. Note that special care must be taken when choosing integration steps and limits, otherwise the demands on computational power can easily skyrocket.

The computational complexity can be partially mitigated by adopting certain simplifications. The vertical profile of the dynamic magnetization (along $z$), for example, largely depends on the exact geometry of the system and should be taken into account for precise calculations. For the purpose of our qualitative estimate of the BLS signal, it can be, however, disregarded. Similarly, the full distribution of the driving field $\boldsymbol{E}_\mathrm{d} (\boldsymbol{r}_{\parallel},z)$ can be, for simplicity, replaced by its mean value over $z$ (possibly augmented by a phase factor due to retardation). Since $\omega \gg \omega_{\mathrm{m} }$, one can also drop the explicit dependence of the driving field on $\omega_{\mathrm{m} }$. These approximations are reflected in the transition between Eqs.~(\ref{Pqvec})~and~(\ref{Dyad}). 
\newline
1) The first key ingredient of our analytical model, the spatial map of the driving electric field $\boldsymbol{E}_\mathrm{d} (\boldsymbol{r})$, is extracted from finite-difference time-domain (FDTD) simulations. Its further processing---such as the calculation of its Fourier transform $\boldsymbol{E}_\mathrm{d} (\boldsymbol{k}_{\mathrm{p} },z)$---is then carried out in Matlab2021a.
\newline
2) The dynamic susceptibility tensor $\hat{\boldsymbol{\chi}} (\omega_{\mathrm{m} },\boldsymbol{k}_{\mathrm{m} })$ acts as the source of inelastic scattering and captures the interaction between spin waves and photons. If we neglect higher order terms like the Cotton–Mouton effect (which are usually small in magnetic metals \cite{hubert2008}), we obtain the following tensor
\begin{equation*}
    \hat{\boldsymbol{\chi}} (\omega_{\mathrm{m} },\boldsymbol{k}_{\mathrm{m} }) = Q_0  \begin{pmatrix}
0 & i m_z & -i m_y\\
-i m_z & 0 & i m_x\\
i m_y & -i m_x & 0
\end{pmatrix},
\end{equation*}
where $Q_0$ is the Voigt constant, and $\left(m_x,m_y,m_z\right)$ are components of a normalized magnetization vector. Since we are not interested in the absolute values but rather in the shape of the signal, we can consider $Q_0=1$. The magnetization vector is obtained from micromagnetic simulations in the same manner as for the phenomenological model (see Extended Data Fig. \ref{Extfig-SimMuMax}).
\newline
3) The dyadic Green's function $\hat{\boldsymbol{G}} (\omega, \boldsymbol{k_\mathrm{p}},\boldsymbol{k_\mathrm{p}}^{\prime})$ is unique for each geometry and it possesses a closed form analytical representation only in a handful of cases. One can, however, reconstruct it approximately from numerical simulations, as the dyadic Green's function can be also viewed as a response of the system to a single plane wave excitation with a propagation vector $\boldsymbol{k_\mathrm{p}}^{\prime}$. The accuracy of this approximation then depends on the number of plane waves used for its reconstruction (in other words the size of the space spanned by $\boldsymbol{k_\mathrm{p}}^{\prime}$). The clear drawback is that each value of $\boldsymbol{k_\mathrm{p}}^{\prime}$ corresponds to one simulation. Since our silicon disks possess a rotational symmetry, the total number of simulations is, fortunately, substantially reduced. Practical implementation of this approach in Lumerical FDTD---although one can employ, in principle, any simulation software---requires the use of a custom source (at least for $k_\mathrm{p}^{\prime} > k_{0}$). The source is inserted into the silicon substrate and its amplitude and spatial profile are set in such a way that the injected wave has the desired $\boldsymbol{E}_{\boldsymbol{k_\mathrm{p}}^{\prime}} e^{i \boldsymbol{k_\mathrm{p}}^{\prime} \cdot \boldsymbol{r} }$ dependence after it enters the space above the permalloy film. The dyadic Green's function $\hat{\boldsymbol{G}} (\boldsymbol{k_\mathrm{p}},\boldsymbol{k_\mathrm{p}}^{\prime})$ can then be identified as the Fourier component $\boldsymbol{E} (\boldsymbol{k_\mathrm{p}})$ of the electric field distribution detected above the silicon disk. 

Besides the parameters and input quantities listed above, there is another important aspect of the analytical model that should be addressed: the order of integration over $\boldsymbol{k_\mathrm{m}}$ with respect to the square modulus operation occurring in Eq.~(\ref{sigma}). It is determined by the coherence properties of the magnon population and in the case of thermal magnons---which are inherently incoherent---the proper procedure is to add intensities of fields derived from magnons with different $\boldsymbol{k_\mathrm{m}}$.

\textbf{Finite-difference-time-domain electromagnetic simulation}
Finite-difference time-domain (FDTD) calculations were performed using Lumerical’s FDTD Solutions software. The 3D simulation region spanned $5.3 \times 5.3 \times 1.36$\,\upmu m$^3$, with the shorter side oriented along the optical axis. Each model included a semi-infinite silicon substrate covered by a 30\,nm thick permalloy thin film on top of which was a 60\,nm thick silicon disk of varying diameter located at the simulation center. Staircase meshing (mesh order 4) was adopted everywhere except in the vicinity of the silicon disk, where it was fixed to 3\,nm cells in all directions. Boundary conditions in the form of perfectly matched layers were used at all simulation boundaries, while the computation was accelerated by application of appropriate symmetry conditions. Gaussian source, implemented using the thin lens approximation, was focused onto the disk-permalloy interface from the air side with the waist diameter set to roughly 600\,nm. The dielectric function of permalloy was taken from \cite{Tikuisis2017} and the dielectric function of silicon was taken from \cite{palik1998}. For the substrate, we used values corresponding to crystalline silicon, while amorphous silicon was assigned to disks to account for the different optical properties of sputtered silicon employed in lithography. The resulting electric field vector components were recorded by field monitors and further processed in Matlab2021a \cite{github}.

\textbf{COMSOL Multiphysics simulation of heat transfer and electromagnetic wave}
The Gaussian beam scattering calculation on the substrate consists of 4 segregated interconnected parts:
\newline 1) Calculation of the Gaussian beam in the air. In the background, a Gaussian beam with E$_x$ polarization is excited. The solution is transferred to the next step using "General Extrusion", where it generates the Gaussian beam at the "in" interface.
\newline 2) Calculation of the Gaussian beam passing through the permalloy layer and the substrate. This wave serves as the background wave in the calculation of Gaussian beam scattering. The electric field strength of the Gaussian beam determined in Part 1 is prescribed at the "in" interface. The "Scattering Boundary Condition" is prescribed at the "out" interface, ensuring that the wave is transmitted out of the domain without reflection. The boundary condition "Perfect Electric Conductor" is prescribed on the vertical boundaries (they are far enough away from the spatially limited Gaussian beam). The solution, extended to the perfectly matched layer (PML) domain (where zero electric intensity is artificially prescribed), is used as the "Background Field" for the next calculation step.
\newline 3) Calculation of the Gaussian beam scattering on the silicon disk on the permalloy layer and the substrate. PML is used to attenuate the scattered wave. "Total Power Dissipation Density" is used as the heat source in the next calculation step.
\newline 4) Calculation of the temperature field in the silicon substrate, permalloy layer and silicon disk. Temperature of 20\,$^\circ$C is prescribed at the "out" interface, the other interfaces are isolated (zero heat flux), and radiation to the surroundings is not considered either. The temperature field is again artificially extended to air and PML, where a temperature of 20\,$^\circ$C is prescribed. The temperature field affects the value of the dielectric function (real and imaginary part) of the silicon.

These 4 steps are solved using the "Frequency-Stationary Solver". Quadratic elements are used in all calculations. The tetrahedral mesh in each domain has a step of $\lambda/5$, where $\lambda$ is the wavelength of the wave in that domain. 

\textbf{Ferromagnetic resonance measurement}
The fundamental mode of the ferromagnetic resonance (FMR) was measured by a broadband ferromagnetic resonance technique using a vector-network analyzer (Rohde\&Schwarz ZVA50). We used so-called flip-chip geometry, where the sample is placed on the excitation antenna with the magnetic layer facing down \cite{Bilzer2007}. We measured the scattering parameter S21 while sweeping the external magnetic field from 0.4 to 0\,T. Then we subtracted the median value of the S21 parameter measured at all fields from the spectra measured at each single field \cite{Vavnatka2021}. In these processed spectra, we were able to find the position of the ferromagnetic resonance by fitting the Lorentzian function, and to obtain the material parameters by fitting the Herring-Kittel formula \cite{Herring1956}.

\textbf{Scattering measurement}
Scattering intensity was measured by a single-particle dark-field confocal spectroscopy. A hollow cone of white light from a halogen lamp was focused onto the sample by a dark-field objective lens (Olympus LMPLANFLN 100$\times$ NA=0.8). The back-scattered light was collected by the same objective lens and then spatially filtered by a 200\,\upmu m pinhole of a multimode optical fiber which subsequently guided the scattering signal to an entrance slit of a spectrometer (Andor Shamrock 303i equipped with an iDus DU420A-BU camera). The relative scattering intensity was calculated as $I_\mathrm{sca} = (I_\mathrm{nd} - I_\mathrm{bg}) / (I_\mathrm{ref} - I_\mathrm{bg})$, where $I_\mathrm{nd}$ is the dark-field signal collected from a silicon disk, $I_\mathrm{bg}$ is the background, and $I_\mathrm{ref}$ is the signal from a spectrally uniform diffuse reflectance standard (Labsphere Inc.).
\section*{Extended Data}
\renewcommand\figurename{Extended Data Fig.}
\begin{figure*}
    \centering
    \includegraphics{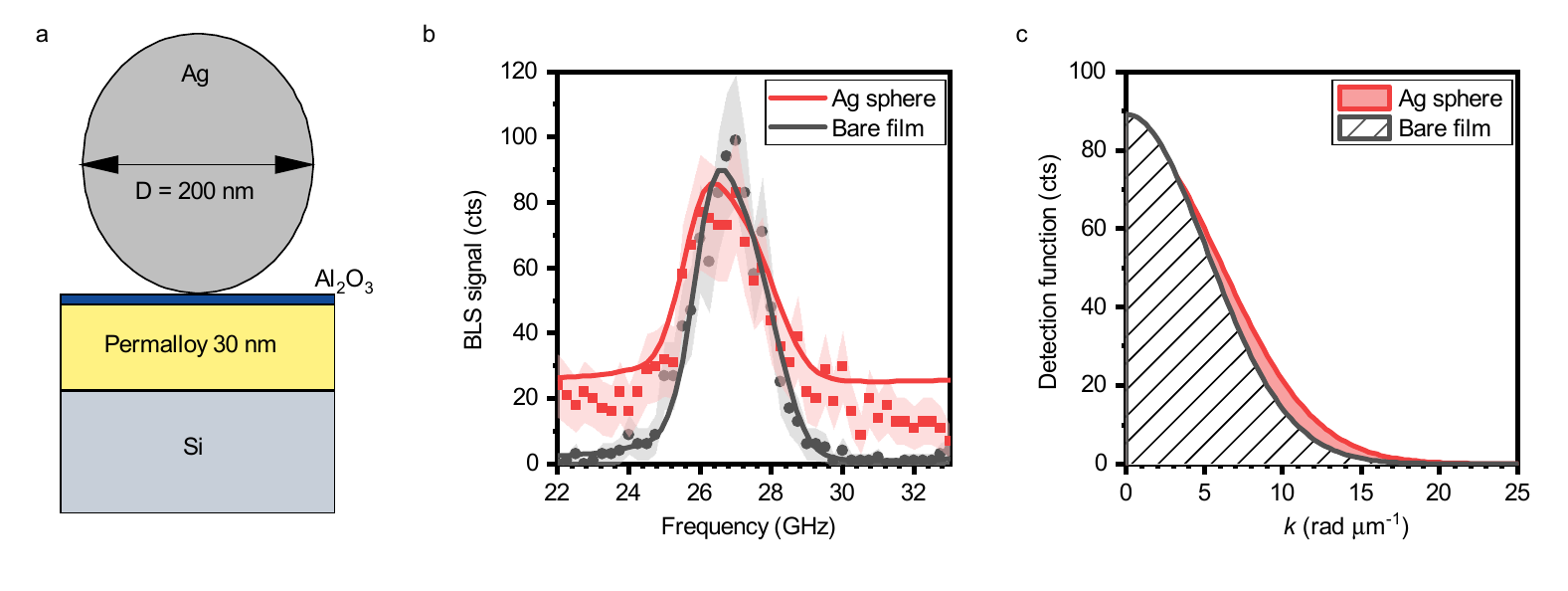}
    \caption{\textbf{Enhancement of the BLS signal by plasmonic nanoparticle.} \textbf{a,} Schematics of a sample configuration used for investigation of plasmon-enhanced-BLS. We used a silver nanosphere with the diameter of 200\,nm, which is tuned to the 532\,nm incident laser light. An insulating Al$_2$O$_3$ spacer layer with the thickness of 2\,nm was used to prevent resonance quenching by the 30\,nm thick permalloy layer underneath. Note that the permalloy layer was the same as the layer used in the experiments described in the main text. \textbf{b,} Thermal spin-wave spectra obtained by \upmu-BLS on the bare permalloy film (black squares) and on the permalloy film with the silver nanosphere (red circles) and their corresponding fits (lines). The error margins represent the 95\% confidence interval. \textbf{c,} Detection function extracted from the fits to the data shown in panel \textbf{(b)}. HWTM of bare film is $11.2 \pm 0.8$\,$\mathrm{rad}$\,\upmu m$^{-1}$ and for silver nanosphere HWTM only slightly increased to $13 \pm 2$\,$\mathrm{rad}$\,\upmu m$^{-1}$.}
    \label{Extfig-Plasmon}
\end{figure*}

\begin{figure*}
    \centering
    \includegraphics{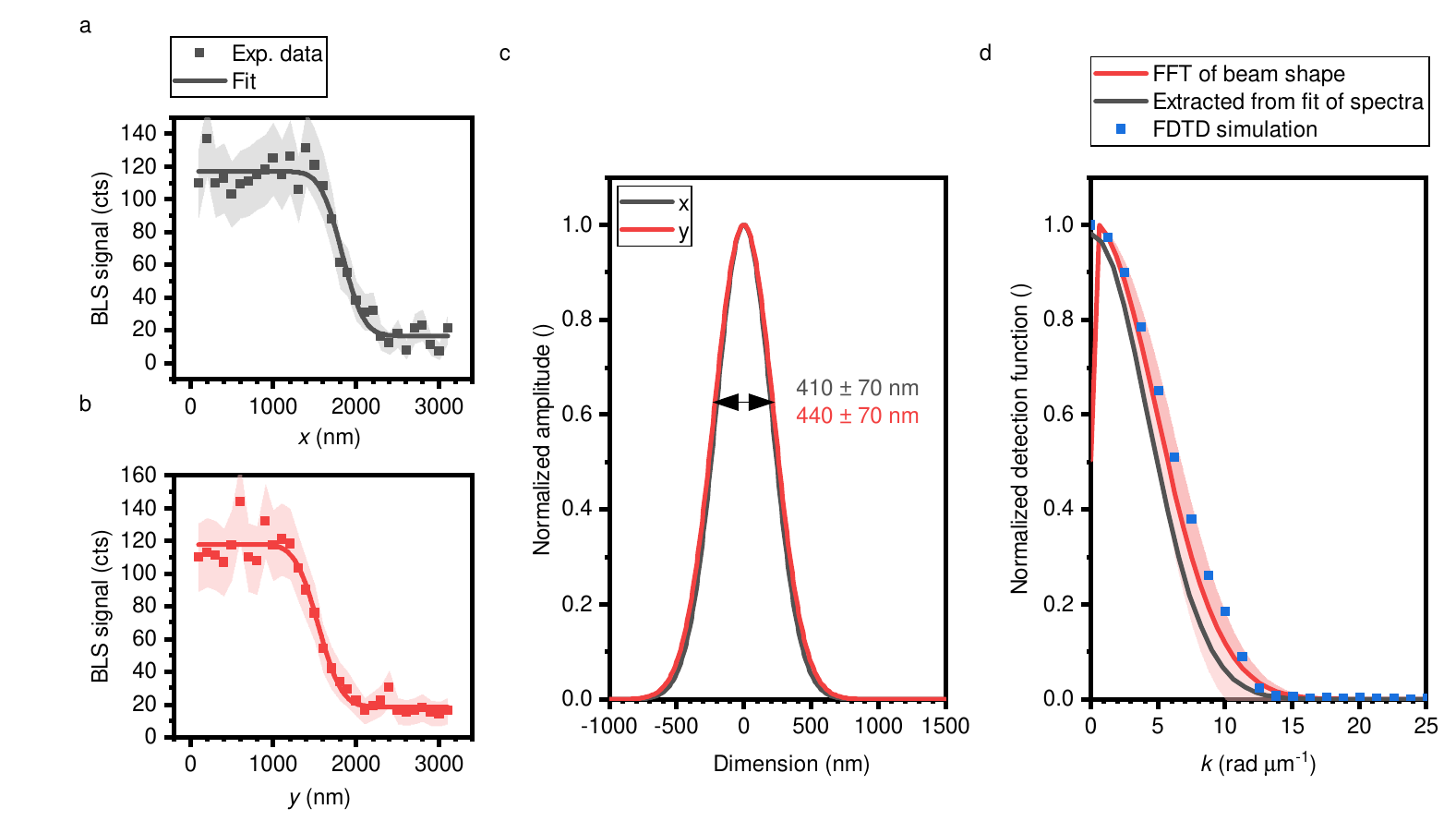}
    \caption{\textbf{Characterization of the Brillouin light scattering microscope beam.}\textbf{ a,b,} The horizontal \textbf{(a)} and vertical \textbf{(b)} knife-edge scans across a magnetic/nonmagnetic boundary.  Black squares are experimental data, solid lines are fits of the error function \cite{flajvsman2016}. \textbf{c,} Beam profiles extracted from the fits in  \textbf{(a)}  (black solid line) and \textbf{(b)} (red solid line). The beam waist sizes obtained from the fits are written next to the data. \textbf{d}, Comparison of the spatial Fourier transform of the knife-edge-measured beam profile (in $x$-direction, red solid line), the detection function obtained from the fitting of the spectra acquired on the bare film (black solid line, see Fig. \ref{fig2}\textbf{c}), and Fourier transform of FDTD simulated beam profile (blue squares). The error margins represent the 95\% confidence interval.}
    \label{Extfig-beamShape}
\end{figure*}

\begin{figure*}
    \centering
    \includegraphics{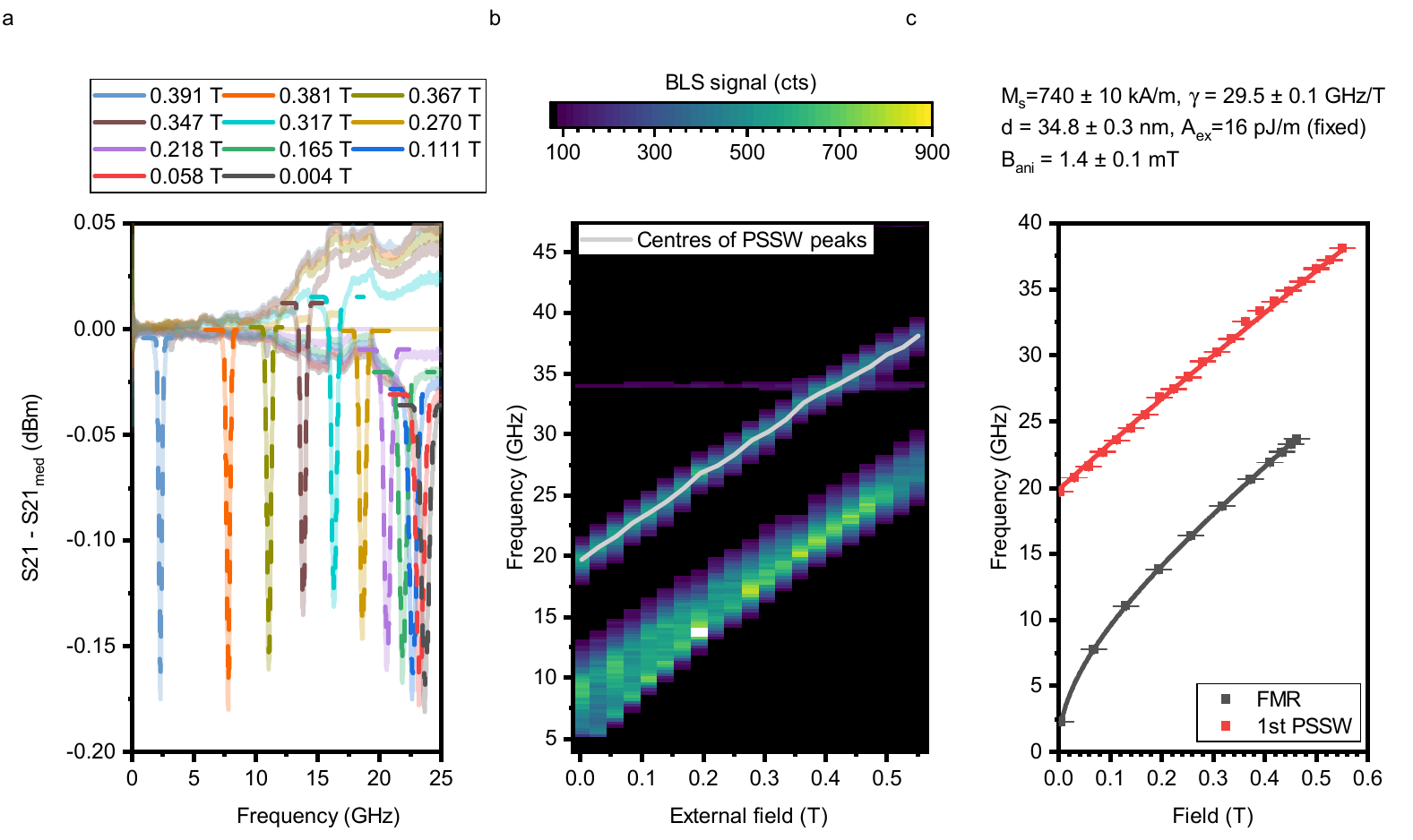}
    \caption{\textbf{Magnetic characterization of the permalloy film used in the experiments.} \textbf{a,} Scattering parameter acquired by a broadband ferromagnetic resonance technique. Different color lines represent different external magnetic fields (see legend). Dashed lines are Lorentzian fits of the resonance peak. \textbf{b,} Thermal magnon spectra measured by \upmu-BLS microscopy. The first perpendicular standing spin wave mode (PSSW) was fitted with Lorentzian function for each frequency bin. The light gray line shows the centers of the Lorentzian functions. The constant signal at approx. 34\,GHz originates from a BLS process on phonons. \textbf{c,} Fits of the data from \textbf{(a)} FMR and \textbf{(b)} PSSW  by the Herring-Kittel formula \cite{Herring1956}. Magnetic properties extracted from the fit are shown above the graph.}
    \label{Extfig-FMR}
\end{figure*}
\newpage

\begin{figure*}
    \centering
    \includegraphics{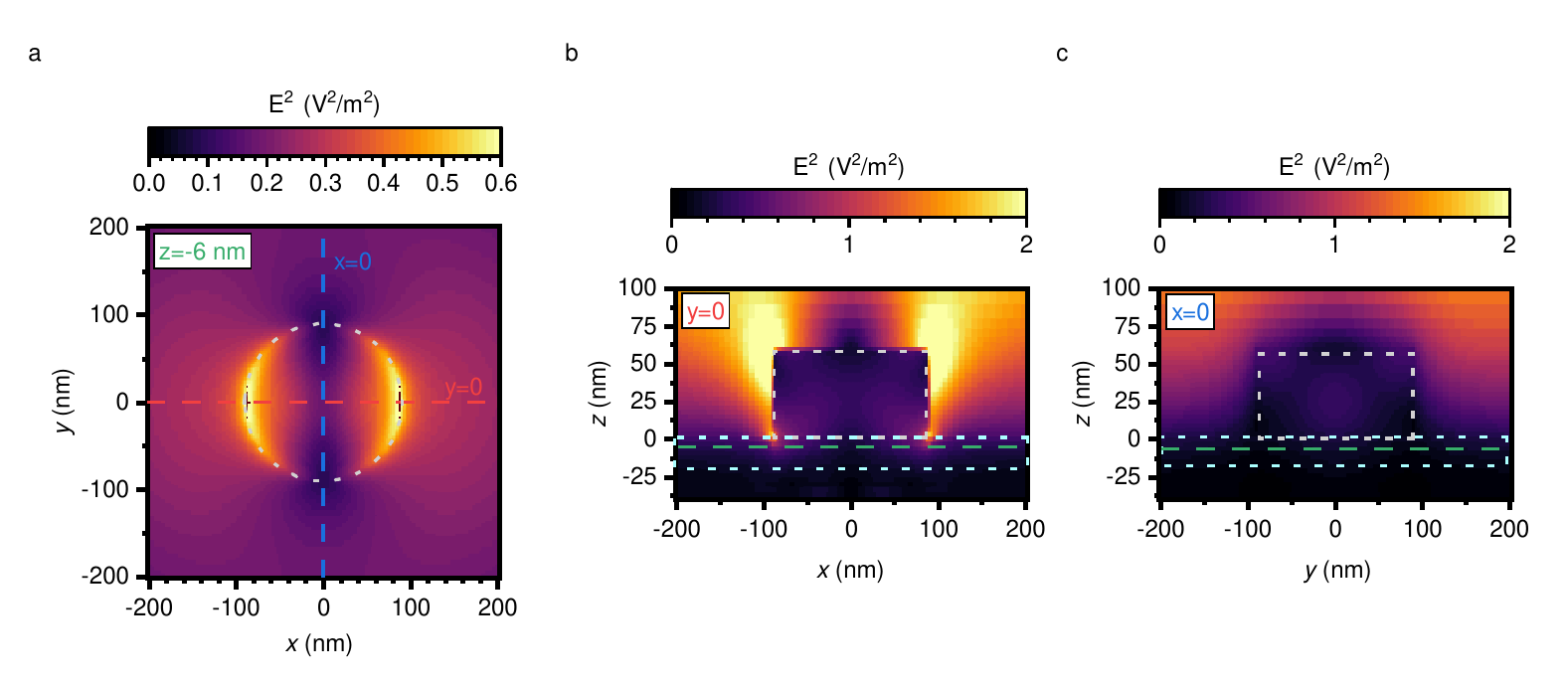}
    \caption{\textbf{Modelled detailed cross-section of 180\,nm wide silicon disk.} \textbf{a,b,c} Squared electric field intensity maps inside a 30\,nm permalloy film. \textbf{a} In-plane cross-sections 6\,nm below the permalloy/air interface.  The two strong hot-spots are formed on the disk boundary (shown as gray dotted line). The red, blue, and green dashed lines in \textbf{a-c)} label \textit{xz}, \textit{yz}, and \textit{xy} cross-sections, respectively. \textbf{b} Cross-section of the \textit{xz} plane. The hot-spots on all edges of the disk are formed. The electric field around the top edge is more intense than around the bottom edge. Nevertheless, it is visible, that significant portion of the light concentrates in the permalloy layer (shown as the cyan dashed rectangle). \textbf{c} Cross-section through \textit{yz} plane. Polarization of the incident Gaussian beam with $\lambda=532$\,nm is in the \textit{x} direction.}
    \label{Extfig-180vs220}
\end{figure*}

\begin{figure*}
    \centering
    \includegraphics{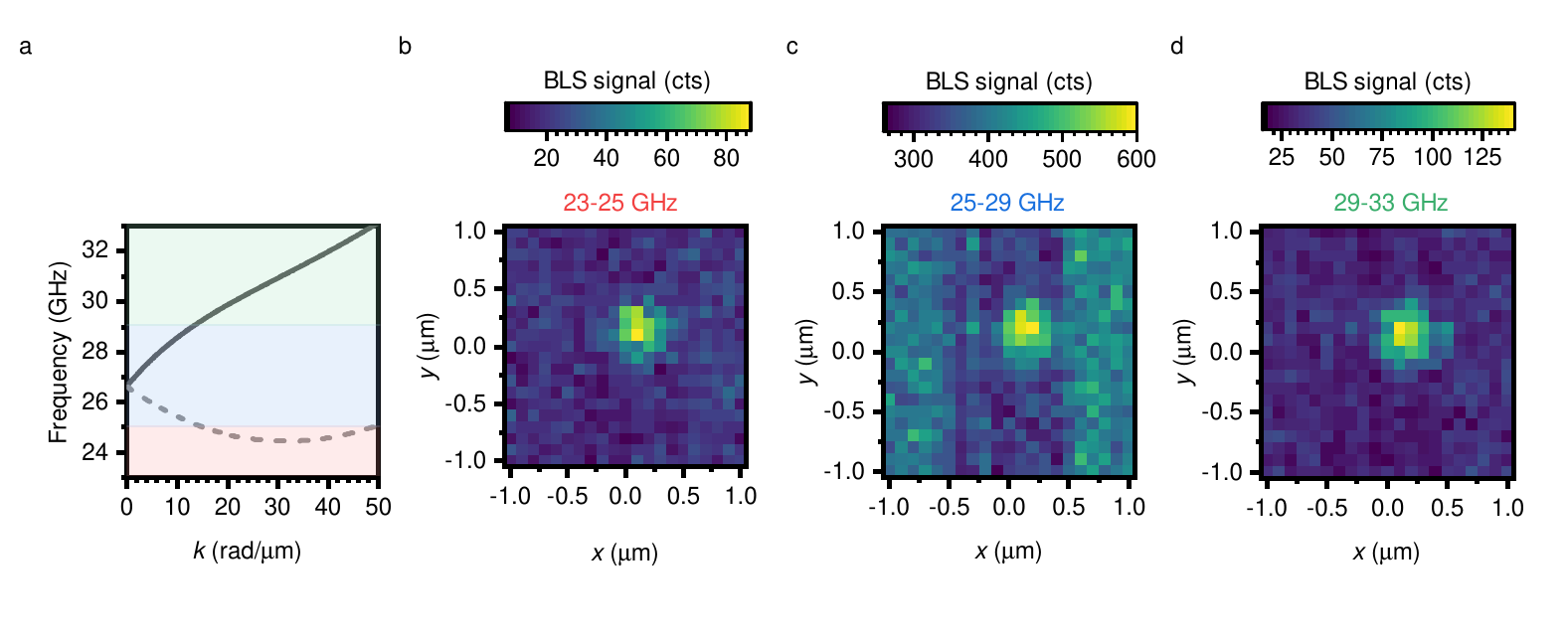}
    \caption{\textbf{Influence of the laser spot position on the BLS signal enhancement.} \textbf{a,} Analytically calculated dispersion relation of the spin waves in a 30\,nm thick permalloy layer. Black solid line represents DE spin waves, gray dashed line represents BV spin waves. The red, blue and green areas show frequency ranges used in panels \textbf{(b), (c),} and \textbf{(d)}. \textbf{b,c,d,}  2D \upmu-BLS maps of (1$\times$1)\upmu m$^2$ area around a 175\,nm wide disk. The BLS thermal signal was integrated in three frequency bands 23-25\,GHz \textbf{(b)}, 25-29\,GHz \textbf{(c)}, and 29-33\,GHz \textbf{(d)}. Data in \textbf{(b)} [\textbf{(d)}] show the region of the dispersion below [above] FMR, which is not achievable with conventional \upmu-BLS. We can see a dramatic increase of the signal in the vicinity of the dielectric nanoresonator. Also note, that the position of the laser beam is crucial, and a movement as small as 100\,nm results in dramatic decrease of the signal. Panel \textbf{(c)} shows the region accessible with conventional \upmu-BLS. Nevertheless, we are able to see an enhancement of the amplitude in the position of the dielectric nanoresonator.}
    \label{Extfig-2Dscan}
\end{figure*}

\begin{figure*}
    \centering
    \includegraphics{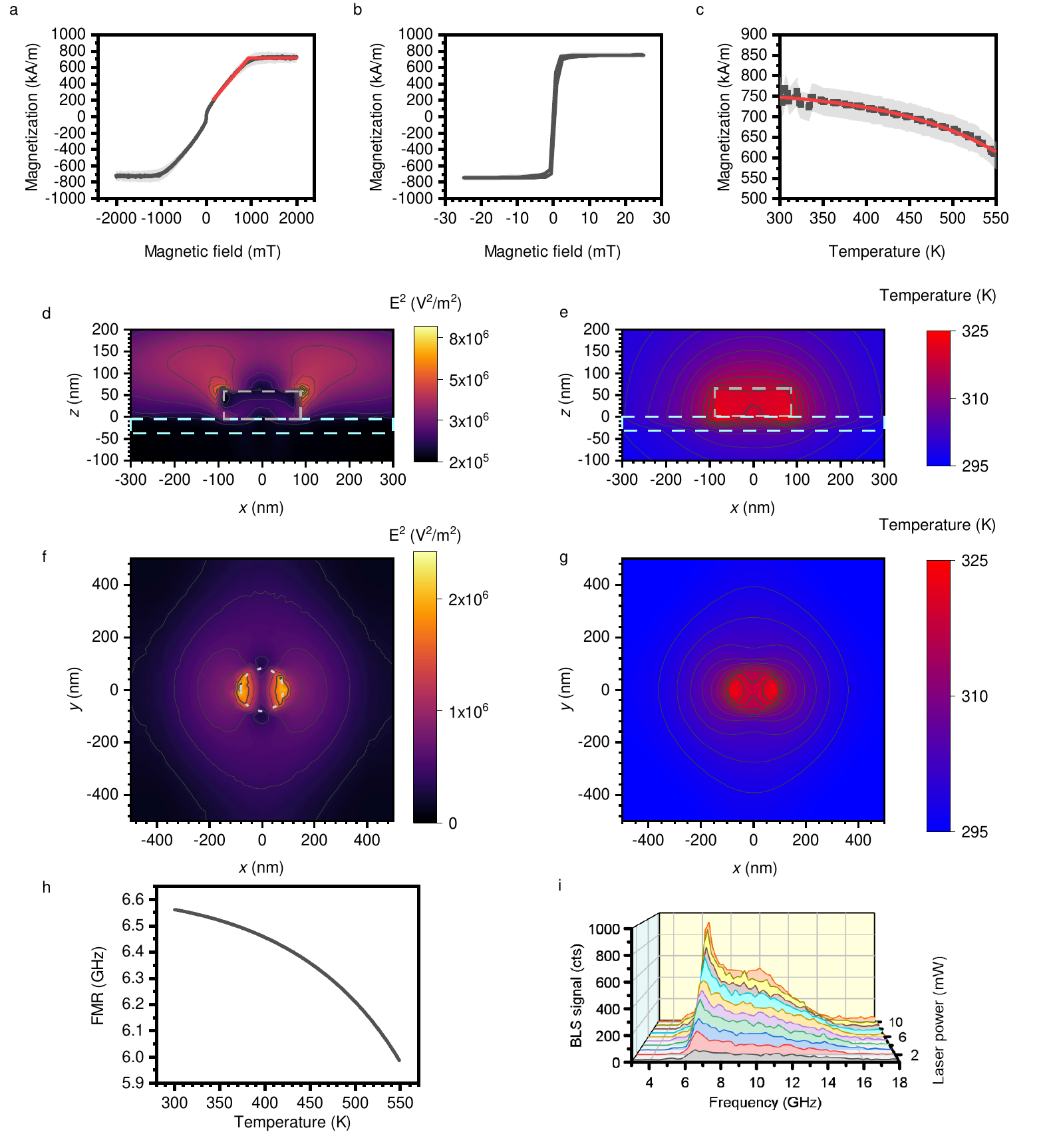}
    \caption{\textbf{Local heating of the sample in the vicinity of the silicon disk.} \textbf{a,} Out-of-plane vibrating sample magnetometry (VSM) measurement of a 20\,nm thick permalloy film (black solid line). The value obtained from the overall magnetic moment and measurement of the volume of the sample is $M_\mathrm{s}=730\pm60$\,kA/m. The red solid line represents a piece-wise fit of the saturating field which gives $M_\mathrm{s}=750\pm10\,\mathrm{kA/m}$. \textbf{b,c} In-plane VSM measurement of the permalloy layer used in our experiments. Temperature dependent measurement of the magnetization in \textbf{(c)} was performed in the field of 20\,mT. The red solid curve is a fit based on the Bloch’s model modified by Kuz’min  $M(T)=M_0\left(1-\left(\frac{T}{T_C}\right)^p\right)^\frac{1}{3}$\cite{Kuzmin2005}. The fitted parameters are $M_0=760\pm30$\,kA/m, $T_\mathrm{C}=670\pm10$\,K, and $p=3.8\pm0.3$. \textbf{d,f,} COMSOL Multiphysics simulations of squared electric field intensity in the \textit{xz} plane \textbf{(d)} and \textit{xy} plane \textbf{(f)}. The results are in qualitative agreement with FDTD simulations reported in the main text. This simulation in addition includes temperature dependent changes in the refractive index of silicon and we see that in this case it does not play any significant role. \textbf{e,g,} The temperature distribution in the \textit{xz} plane \textbf{(e)} and \textit{xy} plane in $z=-6$\,nm \textbf{(g)}. The maximum temperature increase when using the laser power of 3\,mW (the same as in experiment) is 30\,K. \textbf{h,} FMR frequency as a function of temperature calculated from the model presented in \textbf{(c)}. \textbf{i,} BLS spectra measured on a silicon disk with laser power ranging from 1 to 10\,mW. No decrease in FMR frequency is visible even for the highest laser power.}
    \label{Extfig-Temp}
\end{figure*}

\begin{figure*}
    \centering
    \includegraphics{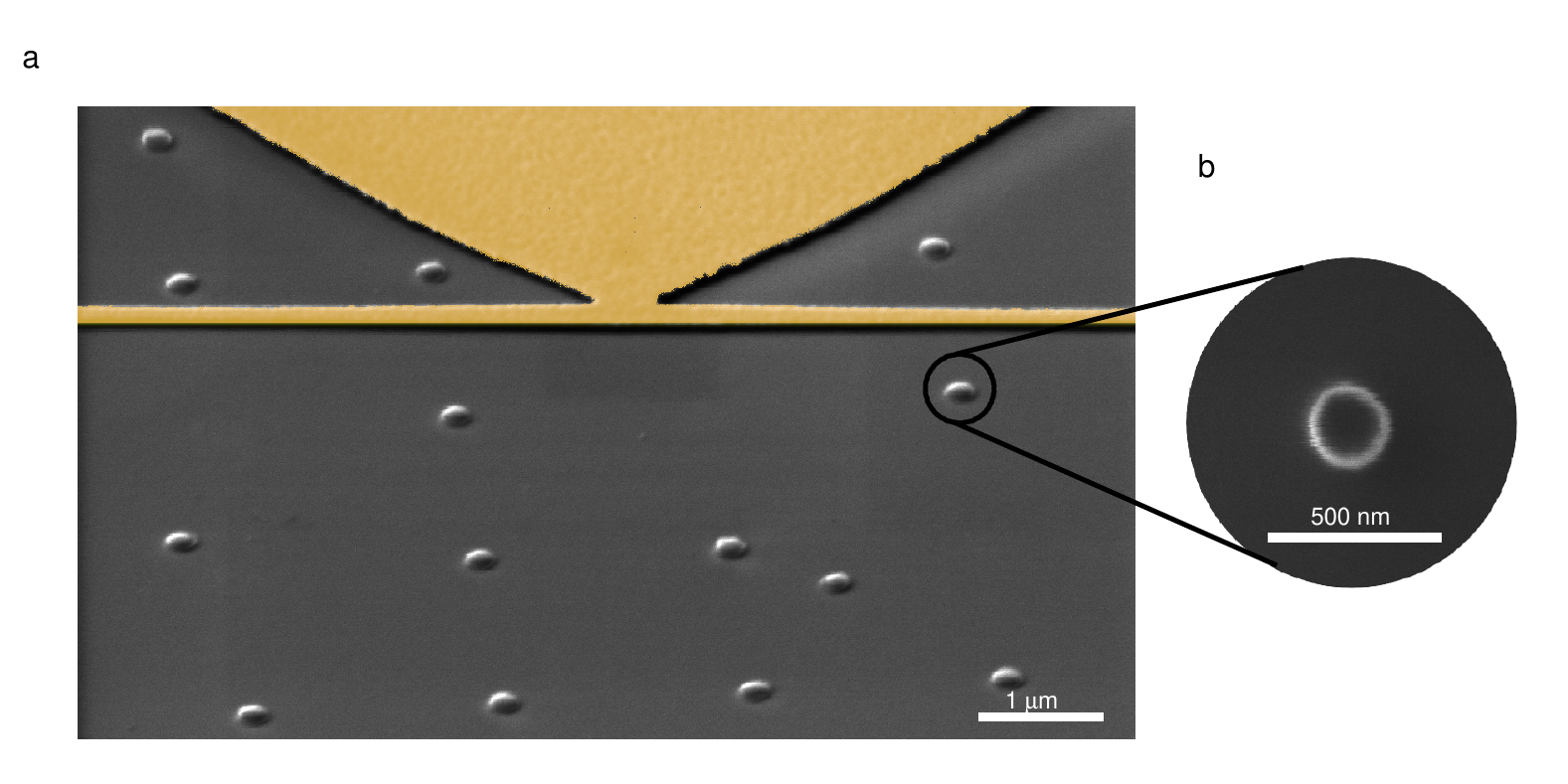}
    \caption{\textbf{Sample for coherent excitation experiments}. \textbf{a,} SEM micrograph of the RF antenna (highlighted by ocher colour) with silicon disks in its vicinity. The image was acquired with the sample tilted by 55\,deg from the surface normal. \textbf{b} Detail of a silicon disk acquired at the tilt of 0\,deg.}
    \label{Extfig-SEM}
\end{figure*}

\begin{figure*}
    \centering
    \includegraphics{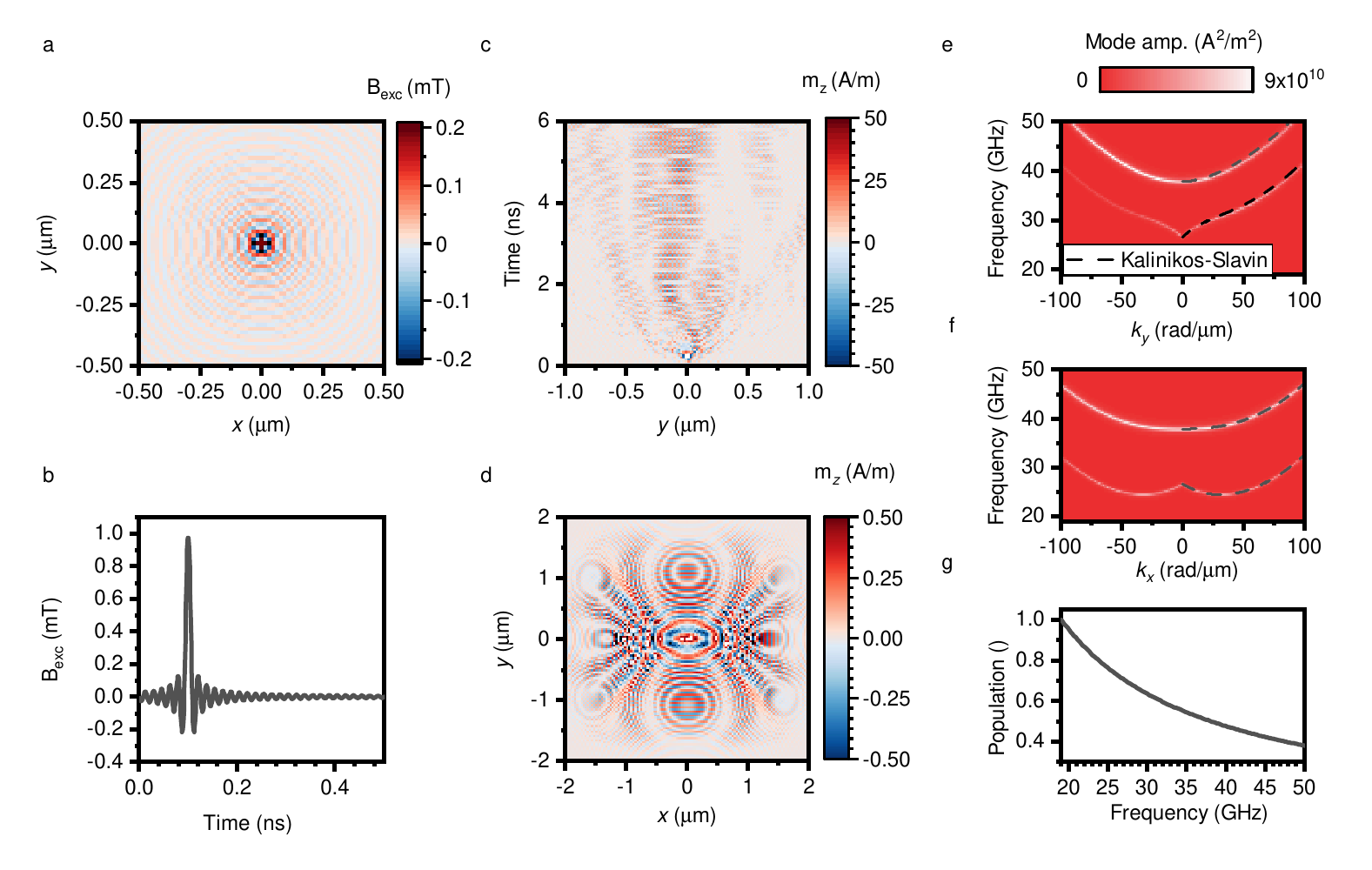}
    \caption{\textbf{Micromagnetic simulations of the spin wave dispersion.} \textbf{a,b }The magnetic moments were excited by 3D sinc function $B_\mathrm{ext}(x,y,t)$.  Panel \textbf{(a)} shows a spatial mask of the external field, while panel \textbf{(b)} shows its time dependence. \textbf{c,} Spatio-temporal evolution of the \textit{z}-component of magnetization at $x=0$. \textbf{d,} Snapshot of the \textit{z}-component of the magnetization at \textit{t}=3\,ns. \textbf{e,f,} Squared Fourier transform of $m_z(x,y,t)$. Panel \textbf{(e)} shows DE spin waves, and panel \textbf{(f)} shows BV spin waves. Dashed lines show the analytically calculated dispersion relations according to the Kalinikos-Slavin model \cite{Kalinikos1986, githubSWT}. \textbf{g,} Normalized Bose-Einstein distribution calculated for a range of frequencies relevant to spin waves in the magnetic field of 550\,mT.}
    \label{Extfig-SimMuMax}
\end{figure*}






\end{appendices}




\end{document}